\documentclass{aa}  

\usepackage{graphicx}
\usepackage{fontawesome5}

\usepackage{algorithm}
\usepackage{appendix}
\usepackage{algorithmic}
\usepackage{xcolor}
\usepackage{caption}
\usepackage{comment}
\usepackage[colorlinks=true,linkcolor=red, citecolor=blue, urlcolor=blue]{hyperref}
\usepackage{txfonts}

\begin{document} 
\count\footins = 1200

\title{The importance of stochasticity in determining  galaxy emissivities and UV LFs during cosmic dawn and reionization}
\titlerunning{Galaxy stochasticity during reionization}

\author{Ivan Nikolić
    \and
    Andrei Mesinger
    \and James E. Davies
    \and David Prelogović
}
\institute{Scuola Normale Superiore, Piazza dei Cavalieri 7, 56125 Pisa, PI, Italy
}

%\date{Received September 15, 1996; accepted March 16, 1997}

% \abstract{}{}{}{}{} 
% 5 {} token are mandatory
 
%\abstract
%{- Investigate the effect of stochasticity at high redshifts for various radiation backgrounds, namely x-rays, Lyman-Werner background and Lyman-continuum photons.}
%{-we develop an analytic formalism that takes into account \textit{all} of the stochastic processes that could contribute at high redshifts. With this formalism we investigate the role of stochasticity for X-ray, LW and LyC emissivities. We investigate the evolution with redshift and assumptions regarding the stochasticity.}
%{Stochasticity in the star-formation that is manifested as a scatter around a SFR-M$_\ast$ relation is crucial to characterize the evolution of all of the radiation backgrounds. Modeling stochasticity becomes more important for higher redshifts going into the Cosmic Dawn. }

\abstract{
 The stochastic nature of star formation and photon propagation in high-redshift galaxies can result in sizable galaxy-to-galaxy scatter in their properties.  Ignoring this scatter by assuming mean quantities can bias estimates of their emissivity and corresponding observables.
 %
 %Here we systematically quantify the relative importance of various sources of stochasticity when computing galaxy ionizing, X-ray and Lyman Werner (LW) emissivities during the Cosmic Dawn (CD) and Reionization (EoR).
 %
 We construct a flexible, semi-empirical model, sampling scatter around the following mean relations: (i) the conditional halo mass function (CHMF); (ii) the stellar-to-halo mass relation (SHMR); (iii) galaxy star formation main sequence (SFMS); (iv) fundamental metallicity relation (FMR); (v) conditional intrinsic luminosity; and (vi) photon escape fraction.
In our fiducial model, ignoring scatter in these galaxy properties overestimates the duration of the EoR, delaying its completion by $\Delta z \sim$ 1--2.
 We quantify the relative importance of each of the above sources of scatter in determining the ionizing, soft-band X-ray and Lyman Werner (LW) emissivities as a function of scale and redshift.
 We find that scatter around the SFMS is important for all bands, especially at the highest redshifts where the emissivity is dominated by the faintest, most "bursty" galaxies.
Ignoring this scatter would underestimate the mean emissivity and its standard deviation computed over 5 cMpc regions by factors of up to $\sim$2--10 at $5\lesssim z \lesssim 15$.
Scatter around the X-ray luminosity to star formation rate and metallicity relation is important for determining X-ray emissivity, accounting for roughly half of its mean and standard deviation.
The importance of scatter in the ionizing escape fraction depends on its functional form, while scatter around the SHMR contributes at the level of $\sim$10--20\%.  Other sources of scatter have a negligible contribution to the emissivities.
 Although scatter does flatten the UV luminosity functions, shifting the bright end by 1--2 magnitudes, the level of scatter in our fiducial model is insufficient to fully explain recent estimates from {\it JWST} photometry (consistent with previous studies).
We conclude that models of the EoR should account for the burstiness of star formation, while models for the cosmic 21cm signal should additionally account for scatter in intrinsic X-ray production.
}

\keywords{Galaxies: high-redshift -- intergalactic medium -- Cosmology: diffuse radiation -- dark ages, reionization, first stars -- X-rays: diffuse background
}

\maketitle
%
%-------------------------------------------------------------------

\section{Introduction}

The Universe underwent dramatic changes during the first billion years. Following cosmic recombination, the Universe was cold, dark and fairly empty.  During the Cosmic Dawn (CD) when the first galaxies formed, their ultraviolet (UV) and X-ray radiation spread out, heating and ionizing the intergalactic medium (IGM).  This culminated in the final major phase change of our Universe: the epoch of reionization (EoR; see for example reviews in \citealt{Zaroubi2013}, \citealt{Mesinger2016} and \citealt{Dayal2018}).

Understanding how the first galaxies heated and ionized the Universe requires modeling their UV and X-ray emission, and constraining these models with data \citep[e.g.][]{Qin2021, HERA2022}.
The emission of any single galaxy is highly variable, depending on the time evolution of star formation, feedback, and geometry of interstellar absorption \citep[e.g.][]{Tacchella2016,Barrow2017, Lovell2021, Pallottini2023}.  
%Indeed highly stochastic star formation in the first simulations has been proposed as one of the explanations of the higher than expected abundance of massive galaxies tentatively detected by JWST at $z>10$ (e.g. Mason+, ..refs).  
These processes are not known from first principles, and are extremely challenging to simulate for a single galaxy, let alone for a cosmological sample of galaxies.

Luckily, the relevant cosmic radiation fields are sourced by the combined contribution from many galaxies, which allows us to take advantage of the Central Limit Theorem and use only {\it average} scaling relations %.  All analytic, semi-numerical and numerical models of the EoR/CD \citep[e.g.][]{Mesinger2011, Holzbauer2011,  Fragos2013, Mirocha2021} rely on such average scaling relations, e.g. in order  to connect galaxy properties to dark matter halos, or to connect unresolved star formation to the properties of the local interstellar medium (ISM). to connect galaxy properties to the dark matter halos that host them.%  The former are poorly known for the high-redshift/faint galaxies  (which are poorly known for the high-redshift/faint galaxies dominating radiation backgrounds) to host dark matter halos (whose abundances and redshift evolution are comperably well known)
to connect galaxy properties to their host dark matter halos (whose abundances and evolution are reasonably well-known).  This is the general approach taken by many analytic, semi-numerical and numerical models of the EoR/CD \citep[e.g.][]{Haiman2000, Ciardi2003, Furlanetto2004,
Mesinger2011, Holzbauer2011, Fragos2013, Ross2017, 
Mirocha2021, Schaeffer2023}.%, with the major differences being in how galaxy properties are parameterized.

However, it is not clear when is it safe to ignore galaxy-to-galaxy scatter (i.e. stochasticity).  Stochasticity can be important even when estimating globally averaged quantities such as the mean EoR history.  Assuming population-averaged quantities (e.g. ionizing escape fraction, stellar to halo mass relation, etc.) can give biased results for correlated distributions (e.g. the average of a product is not the same as the product of the averages; c.f. Appendix \ref{sec:appendix_B} for simple examples).  Moreover, some measurements (e.g. 21-cm interferometry, Ly$\alpha$ forest, kinetic Synaev-Zel'dovich signal, etc.) are sensitive to the spatial fluctuations in the galaxy emissivity, on some range of spatial scales.  As that scale is reduced, there are fewer galaxies over which to average, and stochasticity becomes more important \citep[e.g.][]{Davies2016}.  The importance of stochasticity also increases at high redshifts, where sources are rarer and more biased. It has been evoked to explain controversial claims at $z>10$, such as a rapid redshift evolution of the global 21cm signal during the CD \citep[e.g.][]{Kaurov2018}, and an overabundance of massive galaxy candidates from JWST photometry at $z>10$ \citep[e.g.][]{Mirocha2022, Mason2023, Shen2023}.

Here we construct a model of galaxy emissivity in the bands that are relevant for interpreting current and upcoming observations of the EoR and the CD: (i) ionizing UV (which drives the EoR and determines the residual HI fraction in the ionized IGM); (ii) soft X-ray (which heats and partially ionizes the IGM during the CD); (iii) Lyman Werner (which determines when H$_2$ cooling stops being efficient in the first galaxies).  We compute the distribution of these multi-frequency emissivities as a function of scale and redshift.  Our model samples the largest expected sources of stochasticity, including: 
 the abundance of dark-matter halos, stellar-to-halo mass relation, galaxy main sequence, fundamental metallicity relation, luminosity and escape fraction scalings.  We quantify the relative importance of each term to the total emissivity in each of the considered bands.  We also evaluate the importance of these stochastic terms for simple estimates of the EoR history, as well as the high redshift UV luminosity functions (UV LFs).
Our results can be used to improve estimates of cosmic radiation fields and guide models of the EoR/CD by highlighting the most important sources of scatter.

The structure of this paper is as follows. In Section~\ref{sec:emissivities} we introduce our model for calculating galaxy emissivities. In Section~\ref{sec:results} we present the resulting UV, X-ray, LW emissivity distributions, quantifying the relative importance of each source of stochasticity.  In Section~\ref{sec:EoR} we show two analytic estimates of the EoR history, quantifying the relative impact of ignoring galaxy-to-galaxy scatter.  In Section~\ref{sec:LFs} we show the UV LFs implied by our fiducial model, comparing them to observational estimates from photometric candidates.  Finally, we conclude in Section~\ref{sec:conclusions}. All quantities are presented in comoving units unless stated otherwise. Throughout this work, we assume standard $\Lambda$CDM cosmological parameters ($\Omega_\textrm{m}, \Omega_\textrm{b}, \Omega_{\Lambda}, h, \sigma_8, n_\textrm{s} = 0.310, 0.049, 0.689, 0.677, 0.81, 0.963$), consistent with the latest estimates from \citet{Planck2020}. 

%--------------------------------------------------------------------
\section{Computing emissivities at high redshifts}
\label{sec:emissivities}

If galaxy properties could be written as deterministic functions of the mass of their host halos and/or redshift, we could write the emissivity (e.g. erg s$^{-1}$ cMpc$^{-3}$) in some spectral band, $i$, at a redshift $z$ as:
\begin{equation}
    \varepsilon_i(z) = \int \textrm{d}M_h \frac{\textrm{d}n(M_h, z)}{\textrm{d}M_h} L_i(M_h, z) \ f_{\rm esc, i}(M_h, z) ~ .
    \label{eq:simple-emissivity}
\end{equation}

\noindent Here $\frac{\textrm{d}n}{\textrm{d}M_{\rm h}}$ is the number density of halos per unit mass (i.e. the halo mass function; HMF), $L_i$ is the intrinsic luminosity of a galaxy hosted in a halo of mass $M_h$ at redshift $z$, and $f_{{\rm esc}, i}$ is the fraction of photons  that escape the galaxy to make it into the IGM.

However, we know that the above relations are {\it not} deterministic functions of halo mass and redshift.  Complex physics of galaxy evolution and radiative transfer induces a spread around relations linking different galaxy properties.  Nevertheless, there are empirically well-established relations that characterize some of the main correlations of galaxy properties. Therefore, a more general form for the {\it mean} of the emissivity would marginalize over these relations.  Specifically, we can write the emissivity in a spectral band, $i$, at redshift $z$, averaged over comoving volumes $(4/3) \pi R_{\rm nl}^3$,  as:\footnote{Throughout we use the subscript ''nl'' to indicate non-linear (Eularian) quantities and the subscript ''0'' to indicate Lagrangian quantities linearly evolved to $z=0$ (following convention).  We recall that all length scales are in comoving units, unless otherwise specified.}
%The usual approach in simulations is to let the first term vary as a function of the overdensity of a voxel, but let the other terms be uniquely determined by the mass of the halo. However, complex astrophysics of star-formation and radiative transfer causes both the luminosity and escape fraction to have a spread around the mean. True mean of the emissivity is obtained by marginalizing over various distribution function:

\begin{align}
      \label{eq:first-term}
    \tag{{\rm HMF}}
%    \label{eq:full_emissivity}
  %  \tag{{\rm HMF}}
    \bar{\varepsilon}_i(R_{\rm nl}, z)  = & \int \textrm{d}M_h \int \textrm{d}\delta_0 \frac{\textrm{d}n (M_h, z ~|~R_0, \delta_0)}{\textrm{d}M_h}   p_z(\delta_0 ~|~ R_{\rm nl})  \\
    \tag{{\rm SHMR}}
    \times &\int \textrm{d}M_{\ast} \ p(M_{\ast}~|~M_{\rm h})  \\
    \tag{{\rm SFMS}}
    \times &\int \textrm{dSFR} \ p_z(\textrm{SFR}~|~M_\ast)   \\
    \tag{{\rm FMR}}
    \times &\int \textrm{d}Z \ p_z(Z ~ |~ \textrm{SFR}, M_{\ast}) \  \\
    \tag{{\rm L}}
    \times &\int \ \textrm{d}L_i \ L_i \ p(L_i~|~\textrm{SFR}, Z)  \\
    \label{eq:last-term}
    \tag{{\rm EF}}
    \times &\int \textrm{d}f_{\rm esc, i} \ f_{\rm esc, i} \ p(f_{\rm esc, i})\\
    \label{eq:full_emissivity}
\end{align}
Here $\delta_0(R_0)$ is the linear matter overdensity of a spherical volume of Lagrangian radius $R_0$ corresponding to the final Eulerian radius $R_{\rm nl}$, $p(M_{\ast} ~|~ M_{\rm h})$ is the conditional probability of stellar mass $M_{\ast}$ for a given $M_{\rm h}$, $p_z({\rm SFR}~|~ M_{\ast})$ is the conditional probability of a star-formation rate (SFR) for a given stellar mass\footnote{We use '$z$' subscripts to indicate probability distributions that are also functions of redshift (see below for more details).}, $p_z({\rm Z} ~|~ {\rm SFR}, M_{\ast})$ is the conditional probability of a stellar metallicity $Z$ for a given SFR and $M_{\ast}$, $p(L_i ~|~ \textrm{SFR}, Z)$ is the conditional probability of a luminosity $L_i$ in a given wavelength band $i$ for a given SFR and $Z$, and $p(f_{\rm esc,i})$ is the  probability of an escape fraction $f_{\rm esc,i}$ in band $i$\footnote{Note that our choice of conditional distributions is motivated by well-known relations, but this choice is not unique. Furthermore, we do not assume any direct correlation of the ionizing escape fraction to other properties, as there is currently no consensus on what galaxy property would constitute an appropriate basis to characterize the distribution.}.  Loosely speaking, the running averages of the conditional probabilities in the first four rows are commonly referred to as the halo mass function (HMF)\footnote{Strictly speaking, the average HMF is not the mean of the conditional halo mass function (CHMF), but its limit as $R_0 \rightarrow \infty$ and $\delta_0 \rightarrow 0$.}, stellar-to-halo mass relation (SHMR), star forming main sequence (SFMS) of galaxies, and fundamental metallicity relation (FMR); while (L) and (EF) represent running averages of conditional distributions of the luminosity and escape fraction at wavelength band $i$ , respectively. We label the corresponding rows in the equation above with these acronyms, and go through each probability distribution in more detail below.  In principle, the PDFs above could be conditioned on additional galaxy properties, which could further increase the importance of stochasticity.  Note that higher order moments of the emissivity, such as its variance $\langle (\varepsilon_i - \bar{\varepsilon}_i)^2\rangle$, can be similarly expressed in terms of the above conditional probability distributions.

For general distributions, Eqs.~\ref{eq:simple-emissivity} and~\ref{eq:full_emissivity} do {\it not} give the same mean.  This means that even interpreting average quantities like the EoR history could be biased if not accounting for stochasticity.
More fundamentally, the various sources of scatter in Eq.~\ref{eq:full_emissivity} result in spatial fluctuations in the emissivity which can be important for many EoR/CD observations.
To date, the impact of this scatter on EoR/CD observables has only been explored in a limited fashion.  For example,  \citet{Hassan2021} found that scatter in the intrinsic production rate of ionizing photons predicted by the Simba simulation \citep{Dave2019} has only a modest impact on the EoR morphology.  On the other hand, \citet{Ries2022} used a toy model to characterize the effective scatter in star formation efficiency, finding in some cases a large impact on the EoR and CD morphology; though see \citet{Murmu2023} for the opposite conclusions using a different astrophysical model.  Indeed, a sizable scatter in the ionizing emissivity is needed to explain the latest Lyman alpha forest data at $z=5$ -- 6.3 (\citealt{Qin2021, Gaikwad2023, Davies2024}, Qin et al. in prep.).
% \citet{Kaurov2018} pointed out that in the extreme case that CD galaxies were relatively massive, the IGM heating would proceed very rapidly, potentially being closer to the shape of the putative global signal observed by EDGES (\citet{Bowman2018}, though see \citet{Singh2021}).

%This motivates the calculation of the full distribution of emissivity considering the general shape of Equation~\ref{eq:full_emissivity}. In the case where all of the distributions are normal and scaling relations are simple functions, probability distribution of $\varepsilon$ can be found. In this work we try to be as general as possible.
We could analytically derive the emissivity distribution, $p(\varepsilon_i ~|~ R_{\rm nl}, z)$, if the conditional distributions in Eqs.~\ref{eq:first-term}-\ref{eq:last-term} followed simple Gaussian forms.  In the more general case, we can solve for $p(\varepsilon_i ~|~ R_{\rm nl}, z)$ by numerically sampling the above relations.  
Specifically, to compute a single realization (denoted below with a "tilde") of the emissivity,
%in a region of a given Eulerian scale, $R_{\rm nl}$, at redshift $z$, $\tilde{\varepsilon}_i(R, z)$, 
we perform the following Monte Carlo (MC) procedure:
%\begin{enumerate}[label=\roman*.]
%\item Sample the overdensity, $\tilde{\delta} \sim \mathcal{N}$
%\item Sample the total number of halos...
%\end{enumerate}

\begin{algorithm}
  \caption{Algorithm 1: Computing a single realization of the emissivity in spectral band {\it i} of a region of radius $R_{\rm nl}$ at redshift $z$: $\tilde{\varepsilon}_i(R_{\rm nl}, z)$}
  \label{alg:full}
\begin{algorithmic}[1]
%    \FOR{ some number of iterations}
    \STATE sample the linear matter overdensity $\tilde{\delta}_0 \sim p_z(\delta_0~|~R_{\rm nl})$
    \STATE obtain a realization of the halo field by sampling the CHMF: $\widetilde{\{M_{h}^{\rm j}\}} \sim \textrm{d}n (M_h, z ~|~ \tilde{\delta}_0, R_0)/\textrm{d}M_h$  %\frac{\textrm{d}n}{\textrm{d}M}(\tilde{\delta}, z)$
    \FOR{all halos $j$ with mass $\tilde{M}_{h}^{{\rm j}}$}
    \STATE sample probability that the halo hosts an actively star-forming galaxy, $p(t_{\rm duty}~|~ \tilde{M}_{h}^{\rm j})$
    \hspace{+1cm}\IF {halo does not host a star-forming galaxy} \STATE CONTINUE
    \ENDIF {}    \STATE sample stellar mass $\tilde{M}_{\ast}^j$  $\sim p(M_{\ast}~|~\tilde{M}_{\rm h}^j)$
    \STATE sample star formation rate $\tilde{{\textrm{SFR}}}^j$  $\sim p_z(\textrm{SFR}~|~\tilde{M}_{\ast}^j)$
    \STATE sample metallicity $\tilde{Z}^j$  $\sim p_z(Z~|~\tilde{\textrm{SFR}}^j,\tilde{M}_{\ast}^j)$
    \STATE sample intrinsic luminosity $\tilde{L}_i^j$ $\sim p(L_i~|~\tilde{\textrm{SFR}}^j, \tilde{Z}^j)$
    \STATE sample escape fraction $\tilde{f}_{\rm esc,i}^j$  $\sim p(f_{\rm esc,i})$%~|~\tilde{M}_{\rm h}^j)$
    \ENDFOR
    \STATE $\tilde{\varepsilon_{i}} = \sum_{{\rm halo}=j} \tilde{L}_i^j$ $\tilde{f}_{\rm esc,i}^j$
%    \ENDFOR
\end{algorithmic}
\end{algorithm}

We describe each step of the above MC procedure in turn below.

\subsection{Halo Mass Function (HMF)}
\label{sec:hmf}

In this work we wish to compute the distribution of galaxy emissivities of regions of a given Eulerian scale at a given redshift, $p(\varepsilon_i | R_{\rm nl}, z)$.  Our model is anchored by the fact that galaxies are hosted by dark matter halos, whose relative abundances are described by conditional halo mass functions.

Here we use the hybrid CHMF proposed by \citet{Barkana2004}, in which the analytically-tractable Press-Schecher CHMF \citep{Press1974} is normalized to have the same mean as the (non conditional) Sheth-Tormen HMF \citep[ST,][]{Sheth1999}:
% \begin{equation}
%      \frac{\textrm{d}n}{\textrm{d}M_{\rm h}} (M_{\rm h}, z ~|~ \delta_0, R_{0}) = \frac{\overline{f}_{\rm ST}}{\overline{f}_{\rm PS}} \frac{\overline{\rho}}{M_{\rm h}} \frac{\textrm{d}}{\textrm{d}M} \textrm{erfc} \left[\frac{\delta_{\rm c}(z) - \delta_0}{\sqrt{2\left(\sigma^2(M_h) - \sigma^2(R_{0})\right)}}\right]
%     \label{eq:hmf}
% \end{equation}

\begin{align}
\begin{split}
     &\frac{\textrm{d}n}{\textrm{d}M_{\rm h}} (M_{\rm h}, z ~|~ \delta_0, R_{0}) = \frac{\overline{f}_{\rm ST}}{\overline{f}_{\rm PS}} \sqrt{\frac{2}{\pi}} \frac{\overline{\rho}}{M_{\rm h}}  \times\\& \times\frac{\delta_{\rm c}- \delta_0}{\sigma^2(M_h) - \sigma^2(R_{0})} \left|\frac{\textrm{d}\sigma}{\textrm{d}M_{\rm h}}\right| \exp{\left(-\frac{(\delta_{\rm c}- \delta_0)^2}{\sigma^2(M_h) - \sigma^2(R_{0})}\right)}
     %\frac{\textrm{d}}{\textrm{d}M} \textrm{erfc} \left[\frac{\delta_{\rm c}(z) - \delta_0}{\sqrt{2\left(\sigma^2(M_h) - \sigma^2(R_{0})\right)}}\right]
    \label{eq:hmf}
\end{split}
\end{align}

\noindent In the above, $\overline{f}_{\rm ST}$ and $\overline{f}_{\rm PS}$ correspond to the mean Sheth-Tormen and Press-Schechter collapsed fractions above the atomic cooling threshold of $T_{\rm vir} \geq 10^4$ K, respectively, $\delta_{\rm c}(z)$ is the critical linear density from the spherical collapse model, and $\sigma^2(M)$ is the mass variance of the Lagrangian (linear) density field on scales $M = (4/3) \pi R_0^3$.

In order to sample from Eq. \ref{eq:hmf}, we need to connect Lagrangian and Eulerian quantities (see also, e.g. \citealt{Trapp2020}).  In Lagrangian space, $p(\delta_0)$ follows a zero-mean Gaussian distribution whose width is determined by $\sigma^2(R_0)$.  We transform this distribution to Eulerian space using the spherical collapse model (e.g. \citealt{Mo1996}):
\begin{equation}
    p(\delta_{\rm nl} ~|~ R_{\rm nl}, z) \textrm{d}\delta_{\rm nl} = \frac{1}{1+\delta_{0}} f_{\rm R} (\sigma^2 ~|~ R_0, z) \textrm{d}\sigma^2 ~,
\end{equation}
\begin{multline}
\delta_0 = -1.35 (1+\delta_{\rm nl})^{-2/3} + 0.78785(1+\delta_{\rm nl})^{-0.58661} \\- 1.12431(1+\delta_{\rm nl})^{-1/2} + 1.68647 ~.
\end{multline}
\noindent Here $f_{\rm R}$ is the first-crossing distribution from \citet{Sheth1998}, and $\delta_{\rm nl} = \rho/\bar{\rho} - 1$ is the Eulearian (non-linear) overdensity.

\begin{figure}
    \centering
    \includegraphics[width=\columnwidth]{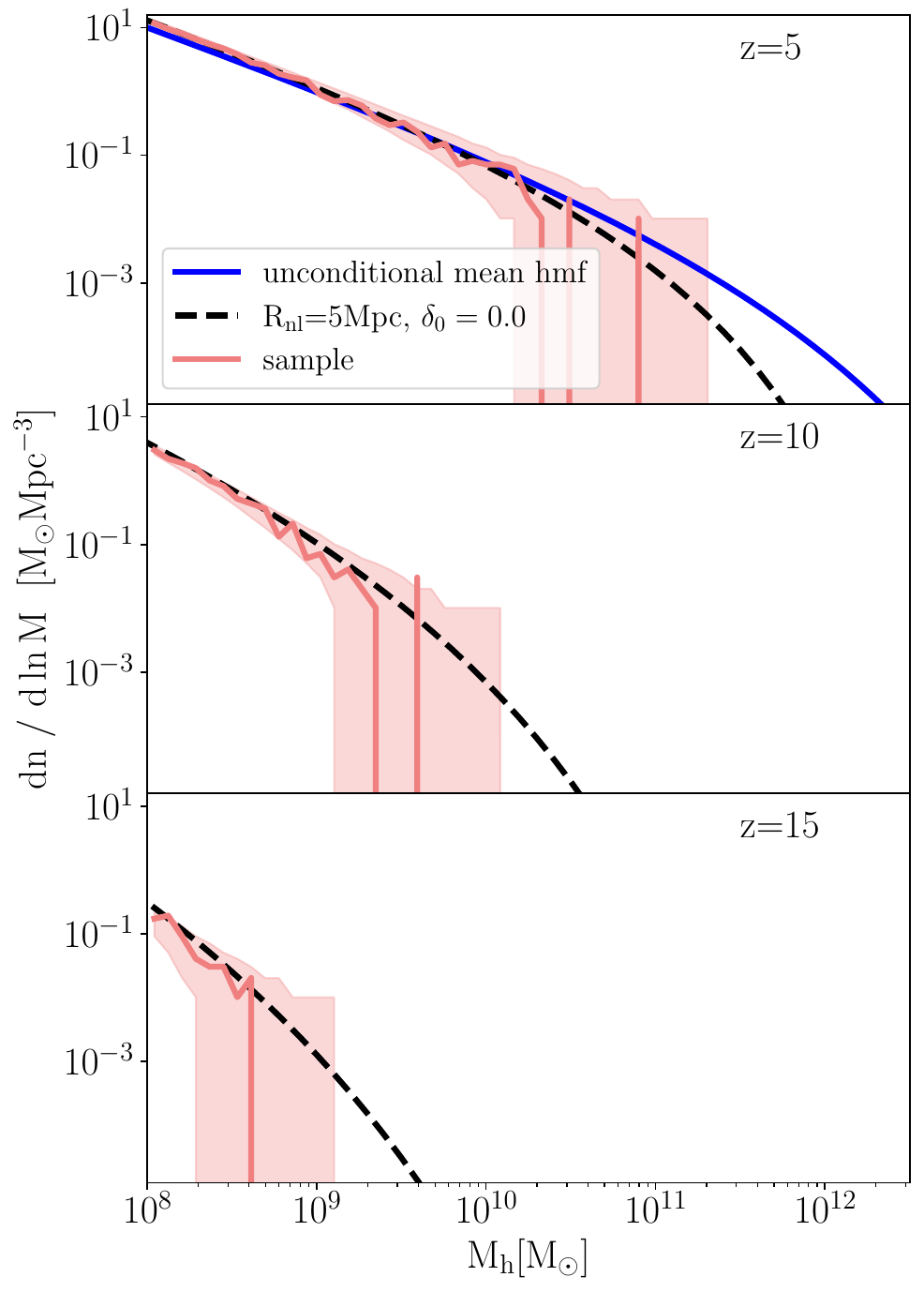}
    \caption{Example halo mass functions used in this work at three different redshifts. The dashed black curves and surrounding red regions correspond to the theoretical mean (Eq.~\ref{eq:hmf}) and 95\% C.L. of the halo field conditioned on a region of scale $R_{\rm nl}=5$ Mpc having a density equal to the cosmic average.  The solid red curve corresponds to a single realization sampled from these distributions. The sample variance scatter in the red curve is seen to increase towards large masses and high redshifts, as the target mean values become smaller.
    In the top panel we also show in blue the non-conditional HMF (i.e. the limit as $R_0 \rightarrow \infty$ at $\delta_0 = 0$).}
    \label{fig:hmfs}
\end{figure}

With the above relations, we generate a Lagrangian overdensity sample, $\tilde{\delta}_0 \sim p(\delta_0~|~R_{\rm nl}, z)$ (step 1 of the MC procedure in the previous subsection).
 We then compute a corresponding realization of the halo field according to the following procedure.
 We first sample the total number of halos with masses above some arbitrary minimum value, obtaining $\tilde{N}(>M_{\rm min} ~|~ \tilde{\delta}_0, R_0)$, by assuming a Poisson distribution whose mean is given by the integral of Eq. \ref{eq:hmf} from $M_{\rm min}$ to infinity.  We then assign each halo a mass by sampling the normalized cumulative mass function, using rejection to ensure the total mass is within $\pm$10\% of the target mean.\footnote{Although approximate, our approach has a couple of notable advantages over other simple MC implementations of stochasticity in which halo numbers are sampled from independent Poisson distributions in fixed mass bins \citep[e.g.][]{Ries2022}.  Firstly, by sampling a continuous CDF, we avoid binning halo masses and forcing them to have discrete values.  Moreover, the mean {\it total} number of halos is much larger than the mean number in any given mass bin, %, i.e. $\bar{N}(>M_{\rm min} | R_0, z) \gg \bar{dN}(>M_h | R_0, z)/d\ln M_h \times d\ln M_h$,
 validating the assumption of a Poisson distribution.  Furthermore, having a mass error threshold ensures approximate mass conservation in each realization.  The alternative of not correlating halo samples in neighboring mass bins and not ensuring mass conservation can significantly overestimate the importance of stochasticity when halos become rare, which can explain why our results are different from those in \citet{Ries2022}.}

We show some example CHMFs in Fig.\ref{fig:hmfs}.  The dashed black curves correspond to the target mean CHMFs in regions of Eularian scale $R_{\rm nl}$, at mean density for infinite realizations, at redshifts $z=5, 10$ and $15$ (top to bottom panels).  The red solid curves show a single realization, computed according to the above procedure, while the red shaded region corresponds to the 95\% C.L. of the halo field conditioned on a region of scale $R_{\rm nl} = 5$Mpc.  The impact of stochasticity in the red curves is very evident as the mean number decreases, i.e. towards high redshifts and high masses.  In the top panel we also show the mean (non conditional)  HMF (i.e. the limit as $R_0 \rightarrow \infty$ at $\delta_0 = 0$).

Fig.\ref{fig:hmfs} also highlights that there are effectively two sources of scatter when determining the halo abundances in a given volume: (i) the scatter in the {\it mean} value of the CHMF, driven by its dependence on the underlying matter overdensity (i.e. the difference between the blue and black curves); and (ii) the scatter due to discrete sampling around the target mean CHMF (i.e. the difference between the black and red curves).  The former determines cosmological signals like 21cm since it is correlated to the underlying matter field.  The latter on the other hand is effectively a sample noise term.
 Both sources of scatter are naturally accounted for in $N$-body simulations, although periodic boundary conditions mean that (i) is underestimated due to limited box sizes (e.g. \citealt{Barkana2004}).  On the other hand, analytic and semi-numerical models of inhomogeneous radiation fields account for (i), but often assume (ii) is negligible in order to reduce computation costs. Below we confirm the validity of this approximation.

\subsection{Stellar-to-halo mass relation (SHMR)}

\begin{figure}
    \centering
    \includegraphics[width=\columnwidth]{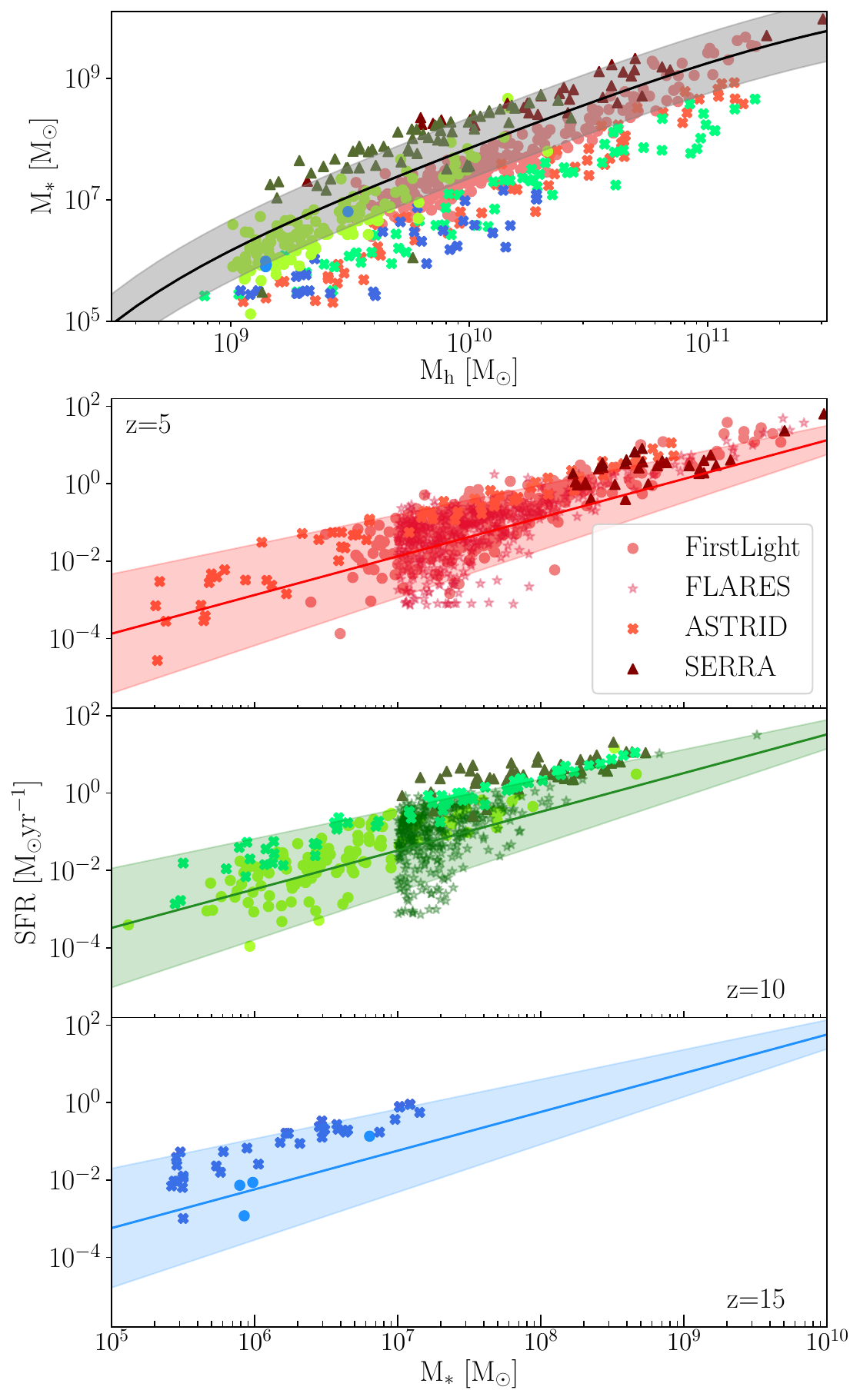}
    \caption{\textit{Uppermost panel:} Our (redshift independent) stellar to halo mass relation ({\it solid curve}) and  2$\sigma$ scatter ({\it shaded region}).  
    \textit{Lower panels:} Galaxy star-forming main-sequence ({\it solid curves}) and 2$\sigma$ scatter ({\it shaded regions}) at $z=5$, 10, 15 ({\it top to bottom}).)  Coloured symbols represent galaxies from cosmological simulations, circles for FirstLight  \citep{Ceverino2018}, stars for FLARES \citep{Lovell2021}, crosses for ASTRID \citep{Bird2022, Davies2023} and triangles for SERRA \citep{Pallottini2022}.  For ASTRID we randomly select galaxies in fixed mass bins, to avoid over crowding the plot, and for SERRA we use their $z=6$ and $z=12$ snapshots for $z=5$ and 10, respectively.}
    \label{fig:SMHM-SFR}
\end{figure}

Both observations and theory established a strong relation between the stellar and halo masses of galaxies \citep[e.g.][]{Harikane2016, Ceverino2018, Stefanon2021, Lovell2021, Kannan2022, Pallottini2022, DiCesare2023}.  %Motivated by results from numerical simulations of high-$z$ galaxies (e.g. \citealt{Ceverino2018, Lovell2021, Bird2022, Pallottini2022}),
Here we assume a log-normal conditional probability of a galaxy having a stellar mass, $M_\ast$, given a host halo mass, $M_h$: $p(\log M_\ast | \log M_h) = \mathcal{N}(\log M_\ast ~|~ \mu_{M_{\ast}}(M_h), \sigma_{M_{\ast}})$.   We assume a mass-independent $\sigma_{M_{\ast}}$ of 0.25 dex \citep[e.g.][]{Ceverino2018, Lovell2021, Pallottini2022} and a mean given by the following double power law SHMR:
%
% \begin{equation}
% \mu_{M_{\ast}}(M_h) = 0.0076 ~ \frac{\left(\frac{2.6 \cdot 10^{11}}{10^{10}}\right)^{0.5}}{\left(\frac{M_{\rm h}}{2.6 \cdot 10^{11} M_{\odot}}\right)^{-0.5} + \left(\frac{M_{\rm h}}{2.6 \cdot 10^{11} M_{\odot}}\right)^{0.6}} ~M_{\rm h}
% \label{eq:SHMR}
% \end{equation}
    
\begin{align}
\begin{split}
    \mu_{M_{\ast}}(M_h) &= \\ -1.412 + & \log{M_{\rm h}} - \log{\left[\left(\frac{M_{\rm h}}{2.6 \times 10^{11} M_{\odot}}\right)^{-0.5} + \left(\frac{M_{\rm h}}{2.6 \times 10^{11} M_{\odot}}\right)^{0.6}\right]} 
    \label{eq:SHMR}
\end{split}
\end{align}

 A standard physical interpretation of the double power law form is that the low-mass scaling is determined by stellar feedback while the high-mass scaling is determined by AGN feedback (e.g. \citealt{Wechsler2018}; \citealt{Behroozi2019} and references therein).  In Eq. (\ref{eq:SHMR}) the normalization and the low-mass power-law index correspond to the maximum a posteriori (MAP) values inferred from a combination of CMB, QSO and high-redshift UV LF observations in \citet{Nikolic2023}, while the high-mass power-law index is taken from the bright-end UV LF empirical fits in \citet{Mirocha2017}.  Our results mostly depend on the former, as the steepness of the HMF at high redshifts means that early radiation fields are dominated by the faint (low mass) galaxies \citep[e.g.][see also below]{Bouwens2015, Bouwens2023, Gillet2020}.
 
Gas accreting from the IGM  onto halos is gravitationally heated, and can also be photo-heated by the ionizing UV background (UVB) during the EoR.  In order to condense onto the galaxy and form stars, this gas needs to cool.  Cooling can be inefficient in halos with small virial temperatures, with an exponentially decreasing fraction of halos capable of sustaining star formation \citep[e.g.][]{Sobacchi2013a, Xu2016}.  Here we account for this effect by assuming only a fraction $f_{\textrm{duty}}(M_{\rm h}) = \exp [-M_{\textrm{turn}}/M_{\rm h}]$ of halos host star-forming galaxies, taking $M_{\rm turn} = 5\times 10^8M_{\odot}$ based on the inference result in \citet{Nikolic2023}.  Specifically, for each halo we sample a random variable uniformly between 0 and 1, and only populate the halo with a star forming galaxy if the value of the random variable is less than $f_{\rm duty}$. 

  In the top panel of Fig. \ref{fig:SMHM-SFR} we show our mean SHMR and 2$\sigma$ scatter (solid black line and gray shaded region, respectively).  The mean SHMR is a power law over most of the mass range shown.  At the high (low) mass end we see a flattening due to our parametrization of AGN feedback (inefficient accretion), as discussed above.     For comparison, we also show galaxies from several hydrodynamic simulations: FirstLight \citep{Ceverino2018}, ASTRID \citep{Bird2022, Davies2023}, and SERRA \citep{Pallottini2022}.  The simulated galaxies are colored according to their redshift, with red for $z=5$, green for $z=10$ and blue for $z=15$.
  
  We see significant differences in the (mean) SHMR between different simulations.  The cosmological zoom-in SERRA simulations imply a mean SHMR that is roughly two orders of magnitude higher at the smallest halo masses compared with the ASTRID simulations.  FLARES and FirstLight are somewhere in between these two extremes, as is our fiducial model.  We remind our reader that our {\it mean} relation was inferred from data, as discussed in \citet{Nikolic2023}, and not based on these simulations.
  
  Conversely, our choice of 0.25 dex scatter around the mean relation is roughly motivated by the galaxy-to-galaxy scatter found in any given hydrodynamic simulation. This scatter is driven primarily by stellar/AGN feedback and mergers.  As our fiducial choice of scatter is motivated by the simulations, we are implicitly including these effects; however, our parametric approach can be used to {\it infer} the mean and scatter in these relations from data in a simulation-agnostic manner. Interestingly, despite the fact that different simulations predict different means, the scatter around the mean is roughly comparable.   Furthermore, we see that the simulations do not show strong evidence of a redshift evolution of the SHMR, justifying our fiducial model \citep[see also, e.g.][]{Mutch2016, Harikane2016, Tacchella2016, Ma2018, Yung2019}.

\subsection{Galaxy star formation main sequence (SFMS)}

The star-formation rates of galaxies, SFRs, are known to be strongly correlated with their stellar mass content.  The mean of this SFR -- $M_{\ast}$ relation is loosely referred to as the galaxy star formation main sequence (SFMS); galaxies with SFRs significantly above (below) the SFMS are referred to as bursty (quenched). The SFMS is well established observationally at low redshifts and (comparably) large masses \citep[e.g.][]{Brinchmann2004, Santini2017, Curtis-Lake2021, Popesso22}. The observed mean relation at small masses follows a power law, whose index is fairly constant but whose normalization decreases with redshift.  This decrease with redshift is naturally reproduced if one assumes that the star formation time-scale is related to the free-fall time at the mean viral density of host halos, $t_{\rm ff}$, which during matter domination scales as the Hubble time: $t_{\rm ff}\propto H^{-1}(z)$ \citep[e.g. see][and references therein]{Park2019}.  

%Cosmological simulations reproduce the observed SFMS (indeed their sub-grid feedback models are calibrated to do so); however, it is not clear how this relation extends to the low masses and high redshifts where we have no observations.  More fundamentally, the level of scatter and how this scatter depends on redshift and mass is poorly known \citep[e.g.][]{Ceverino2018, Lovell2021, Davies2022}. 

At a given $z$, we again assume a log-normal conditional probability $p_z(\log {\rm SFR} ~|~ \log M_\ast) = \mathcal{N}[\log {\rm SFR} ~|~ \mu_{\rm SFR}(M_\ast, z), \sigma_{{\rm SFR}}(M_\ast)]$.  For the mean SFMS we use the model of \citet{Park2019}, with the normalization set by the MAP values in \citet{Nikolic2023}:
\begin{equation}
    \mu_{\rm SFR}(M_{\ast}, z) = \log{M_{\ast}} - \log{[0.43 H^{-1}(z)]}
\label{eq:SFR}
\end{equation}
We assume a mass-dependent scatter that increases towards smaller masses, as these galaxies are expected to be more bursty\footnote{Throughout this work we use "burstiness" to indicate a wide scatter around the mean SFMS. We do not investigate what such distributions imply for the star formation histories of individual galaxies.}:
\begin{equation}
    \sigma_{\rm SFR}(M_{\ast}) = 
\begin{cases}
    -0.12  \log{M_{\ast}} + 1.35 &\text{if $\log M_{\ast}<10$} \\
    0.19 & \text{otherwise}
\end{cases}
\label{eq:sfr-scatter}
\end{equation}
The normalization and scaling of the scatter was fit to the hydrodynamic simulations of \citet{Ceverino2018}, but the mean (i.e., Eq.~\ref{eq:SFR}) is the same as the one in \citet{Nikolic2023}. Just like with the SHMR, the scatter is mostly driven by galactic feedback and mergers.

We plot our assumed SFMS and 2$\sigma$ scatter in the bottom three panels of Figure~\ref{fig:SMHM-SFR}.  Panels correspond to $z=5$, 10, 15 ({\it top to bottom}), with different symbols indicating values taken from hydrodynamic simulations: FirstLight \citep{Ceverino2018}, FLARES \citep{Lovell2021}, ASTRID \citep{Bird2022}, and SERRA \citep{Pallottini2022}.  The figure illustrates that our fiducial model is in general agreement with results from these hydrodynamic simulations.\footnote{Detailed comparisons to other works would require standardizing definitions.  For example, here we define the SFR as an instantaneous quantity, while elsewhere it could be averaged over $\sim100$ Myr to allow for a more direct comparison to photometric observations.  Here we are just interested in confirming that our fiducial choices are reasonable.}

\subsection{Fundamental Metallicity Relation (FMR)}
\label{sec:metalicity}

\begin{figure}
    \centering
    \includegraphics[width=\columnwidth]{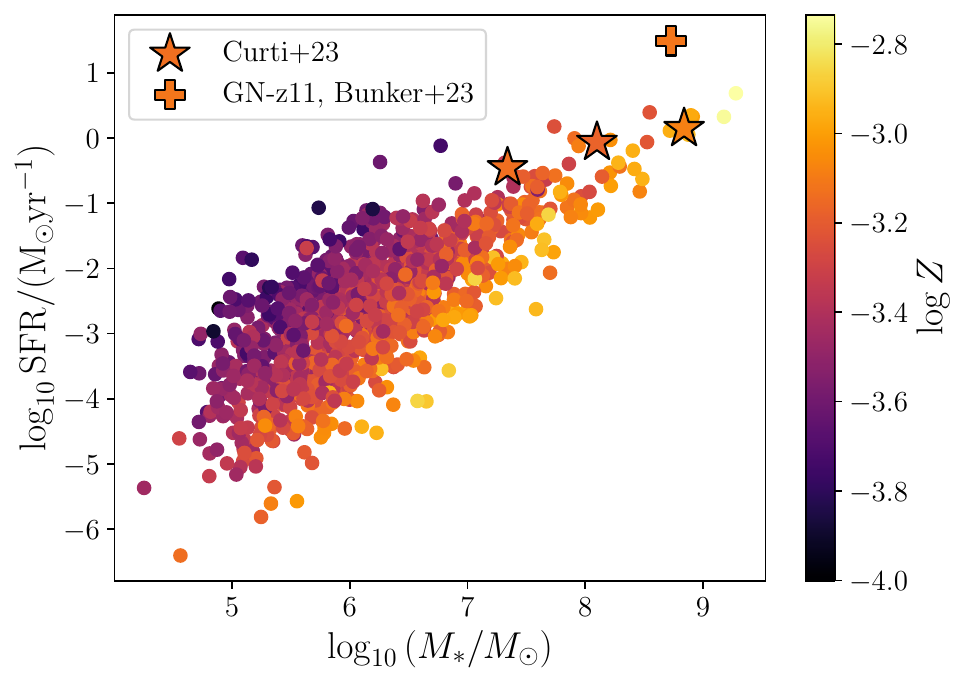}
    \caption{The stellar metalicities, stellar masses and star formation rates of galaxies from a single realization of a $R_{\rm nl} = 5$ cMpc volume at mean density at $z=6$.  Each point corresponds to a single galaxy, with the color denoting its metallicity: $\tilde{Z}^j$  $\sim p_z(Z~|~\tilde{\textrm{SFR}}^j,\tilde{M}_{\ast}^j)$ (see text for details). Stars denote the observationally-estimated means in three bins for $z>6$ galaxies in \citet{Curti2023} (converted from gas metalicities, see text for details), as well as the metallicity estimate of GN-z11 from \citet{Bunker2023}.}
    \label{fig:metalicity_plot}
\end{figure}

The galaxy emissivity also depends on the metallicity of the stellar population. Here we relate the metallicity to the SFR and stellar mass of a galaxy, taking advantage of the well-studied fundamental metallicity relation (FMR; \citealt{Mannucci2010, Curti2020}). Specifically, we assume a log-normal conditional probability of a galaxy having a stellar metallicity $Z$, given its SFR and stellar mass, $p_z(\log Z ~|~ \log {\rm SFR}, \log M_\ast) = \mathcal{N}[\log Z ~|~ \mu_Z(M_\ast, {\rm SFR}, z), \sigma_Z]$. We assume a constant scatter of $\sigma_Z = 0.1$ dex, and a mean given by the following (c.f. \citealt{Curti2020}):

% The galaxy emissivity also depends on the metallicity of the stellar population.
% %Amount of metals in a galaxy strongly affects the evolution of the ISM as well as the spectral energy density (SED) of the galaxy. As the galaxies grow, stars pollute the ISM through outflows and SNe. This processes link the metalicity to the stellar mass of the galaxy. Observations also point to a decrease in metalicity with the increased SFR for a fixed stellar mass. This two scaling give way to a 
% Here we relate the metallicity to the SFR and stellar mass of a galaxy, taking advantage of the well-studied fundamental metalicity relation (FMR) \citep{Mannucci2010, Curti2020}. Specifically, we assume a log-normal conditional probability of a galaxy having a stellar metalicity $Z$, given its SFR and stellar mass, $p_z(\log Z | \log SFR, \log M_\ast) = \mathcal{N}[\log Z ~|~ \mu_Z(M_\ast, {\rm SFR}, z), \sigma_Z]$.  We assume a constant $\sigma_Z=0.1$ dex in Z around the mean FMR:
% \begin{align}
%    \mu_Z((M_\ast, {\rm SFR}, z) &= \\
% \noindent   &12 + \log{(\textrm{O}/\textrm{H})} \ (M_{\ast}, \textrm{SFR}) = 8.799 - \\ &\frac{0.31}{2.1} \log{\left(1 + \left(\frac{M_{\ast}}{M_0(\textrm{SFR})}\right)^{-2.1}\right)}
%     \label{eq:metalicity_eq}
% \end{align}

\begin{equation}
    \mu_{Z} (M_\ast, {\rm SFR}, z) = 
    0.296 \left( 1+ \left( \frac{M_{\ast}}{M_0}\right)^{-2.1} \right)^{-0.148} 10^{\Delta_z} \ Z_{\odot} ~, 
\end{equation}

\noindent where $M_0 (\rm SFR) \equiv 10^{10.11} \times ({\rm SFR / M_{\odot}{yr^{-1
}})}^{0.56} M_{\odot}$, and $\Delta_z = -0.056 z + 0.064$ accounting for putative redshift evolution \citep{Curti2023}.  In the above, we converted from gas phase to stellar metalicities using $Z/Z_{\odot} = 10^{(12+\log{(\textrm{O}/\textrm{H})} - 8.69)}$ with solar metallicity $Z_{\odot} = 0.02$ \citep{Asplund2004}, and adjusting for gas phase metallicities being higher by a factor of $\approx 2.63$ on average \citep{Strom2018}.  

%with $\log (M_0(\textrm{SFR})) = 10.11 + 0.56 \log{\textrm{SFR}}$. We convert gas-phase metalicities to stellar ones by using $Z/Z_{\odot} = 10^{(12+\log{(\textrm{O}/\textrm{H})} - 8.69)}$ with solar metalicity $Z_{\odot} = 0.02$ \citep{Asplund2004}.

%There is some tentative theoretical (e.g. \citealt{Dave2011, Torrey2019, Garcia2024}) and % FMR is established as an equilibrium between the metal production from the stars and dilution due to the inflow of pristine gas. At higher redshifts, gas fraction is higher leading to  lower metalicities for a given stellar mass \citep{Finlator2008, Dave2011}. 
%observational (e.g. \citealt{Langeroodi2023,Curti2023, Nakajima2023}) evidence to suggest that the mean FMR decreases with increasing redshift. Here we use the relation from \citet{Curti2023}: $\Delta_z = -0.056 z + 0.064$, to account for this putative evolution.

In Fig.~\ref{fig:metalicity_plot} we show galaxies from a single realization of a comoving volume with radius $R_{\rm nl} = 5$ cMpc, at mean density at redshift $6$. Each point denotes a single galaxy with the color corresponding to its typical stellar metallicity.  Note that the apparent scatter in the metallicity at a fixed $M_\ast$ is considerably larger than the 0.1 dex scatter we set around the mean FMR at a given $M_\ast$ and SFR.  This is because the intrinsic scatter in the combination of SHMR and SFMS (i.e. the width of $p_z({\rm SFR}, M_\ast)$ from the previous sections) dominates over our choice of scatter in the metallicity {\it given} these properties (i.e. the width of $p_z(Z ~|~ {\rm SFR}, M_\ast)$; see also e.g. \citealt{Garcia2024}).  In the figure we also show the binned values for the metallicity of $z>6$ galaxies from \citet{Curti2023} as well as the metallicity estimate of GN-z11 from \citet{Bunker2023}.  Although these observations span a range of redshifts, they are generally consistent with our samples.

\subsection{Luminosity scalings}

The intrinsic luminosity of a galaxy depends primarily on the SFR and its history, as well as the metallicity of the stellar population \citep[e.g.][]{Brammer2008, AllendePrieto2018, Stanway2018, Lehmer2021, Fragos2023}. Here we describe how we compute the intrinsic luminosities for each of the wavelength bands of interest: X-ray, ionizing and Lyman Werner.

\subsubsection{Soft-band X-ray luminosity}

\begin{figure}
    \centering
    \includegraphics[width=\columnwidth]{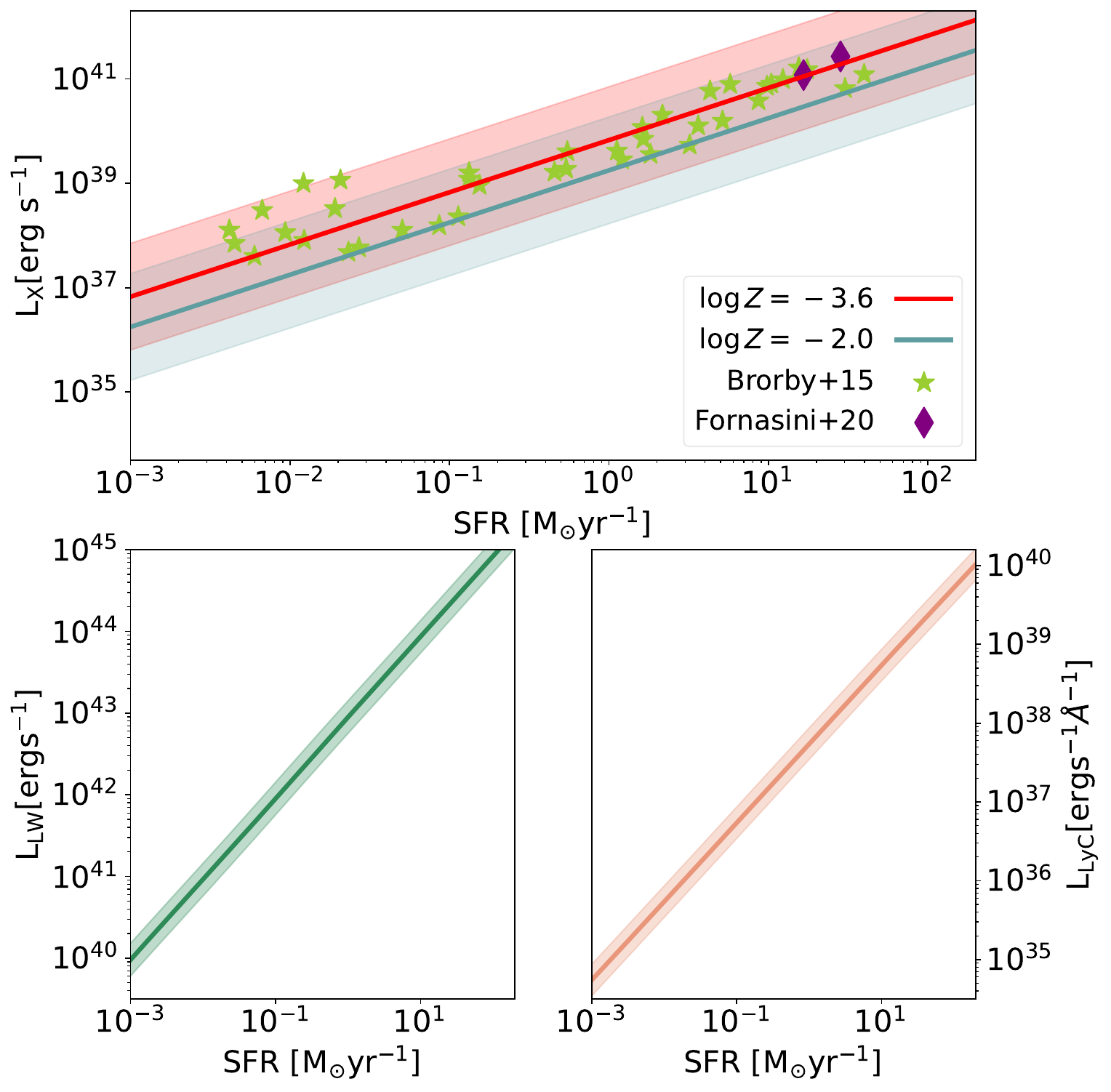}
    \caption{\textit{Upper panel:} scaling of soft-band, high-mass X-ray binary luminosity with SFR. Red and blue lines with the corresponding shaded regions represent the mean and $2\sigma$ range for metalicities $Z=-3.6, -2.0$, respectively.  Green stars correspond to values from local star forming galaxies, discussed in \citet{Brorby2016}. Purple diamonds are redshift-binned stacks from \citet{Fornasini2019} for $z = 1.5$ and $2.3$.  \textit{Lower panels:} SFR scaling of the integrated Lyman-Werner (11.2 -- 13.6 eV) luminosity (\textit{left panel}) and specific ionizing luminosity at the Lyman-limit (\textit{right panel}). Shaded regions represent the $2\sigma$ scatter around the mean relation. Both panels assume that metallicity follows the mean FMR.}
    \label{fig:Lx_SFR}
\end{figure}

Soft-band\footnote{Here we define the soft band to be $0.5-2 \ \textrm{keV}$. Roughly speaking, photons with higher energies do not interact with the high-$z$ IGM \citep[e.g.][]{Peng2001, Xu2014, Madau2017}, while photons with lower energies get absorbed inside the host galaxies \citep[e.g.][see also Section~\ref{sec:escape_fraction}]{Das2017}.} X-rays emerging from the first galaxies are responsible for heating and partially ionizing the IGM during the cosmic dawn \citep[e.g.][]{McQuinn2016}, which can have a dramatic imprint in the cosmic 21cm signal \citep[e.g.][]{Mesinger2013, HERA2022}.  It is likely that the X-ray emissivity of $z>6$ galaxies is dominated by high mass X-ray binaries \citep[HMXBs; e.g.][]{Furlanetto2006, Fragos2013, Pacucci2014, Eide2018}.  HMXBs are massive stars accreting onto a compact companion.  The total X-ray output of a galaxy from HMXBs should therefore scale with the SFR of the galaxy (due to the rapid stellar evolution timescales of massive stars) and its metallicity (which determines the efficiency of radiative-driven winds and the resulting mass loss of the massive companion).  Indeed we observe a strong dependence of the X-ray luminosity on the galaxy's SFR and metallicity in local galaxies and in stacks out to $z\sim2.5$ \citep[e.g.][]{Brorby2016, Lehmer2016, Fornasini2019, Lehmer2021}.

Here we assume a log-normal conditional probability of a galaxy having an intrinsic soft-band X-ray luminosity, $L_{\rm X}$ (in units of erg s$^{-1}$), given a SFR and metallicity: $p(\log L_{\rm X} ~|~ \log {\rm SFR}, \log Z) = \mathcal{N}(\log L_{\rm X} ~|~ \mu_{\rm X}({\rm SFR}, Z), \sigma_{\rm X})$.   We assume a constant $\sigma_{\rm X}$ of 0.5 dex and a mean given by:
%
% \begin{align}
% \begin{split}
%     \mu_{\rm X}({\rm SFR}&, 12+\log{(\textrm{O}/\textrm{H})}) =\\& \quad 
%     (-0.11 \times (12+\log{(\textrm{O}/\textrm{H})})+2.30) \ \log{\textrm{SFR}}\\&+(-0.31 \times (  12+\log{(\textrm{O}/\textrm{H})})+41.78) ~ .
%     \label{eq:Lx-sfr}
% \end{split}
% \end{align}
\begin{align}
\begin{split}
    \mu_{\rm X}({\rm SFR}, Z) = 
    \log{\textrm{SFR}}+40.5 + \log{\left[\left(\frac{Z/Z_{\odot}}{0.05}\right)^{0.64}+1\right]}.
    \label{eq:Lx-sfr}
\end{split}
\end{align}
These fiducial choices are obtained by assuming a double power-law function for the X-ray luminosity function that results in a flattening at lower metallicities and fits the data at the high SFR/metallicity end \citep[e.g.][]{Fragos2013, Lehmer2021, Kaur2022, Geda2024}. They are roughly consistent with empirical fits to local galaxies, which could however suffer from incompleteness at the lowest SFR bins (e.g. \citealt{Brorby2016}).  We have converted the hard band X-ray luminosity in \citet{Lehmer2021} ($0.5$-$8$ keV) to the soft band one by multiplying the luminosities by a factor of $0.3$, consistent with observational estimates \citep[e.g.][]{BasuZych2013} and corresponding to an intrinsic SED with a power-law index of $\Gamma = 2.0$ \citep[e.g.][]{Mineo2012}.

We show this dependence of the X-ray luminosity with SFR, for two different metallicity values, in the top panel of  Figure~\ref{fig:Lx_SFR}. For comparison, we include redshift-binned stacked observations of star-forming galaxies at $z\sim2$ from \citet{Fornasini2019} and Lyman-Break analogues from \citet{Brorby2016}.  We see that our fiducial model is consistent with current data; however, it is highly uncertain how these relations scale to the first galaxies whose metallicity ranges are not sampled by current observations \citep[e.g.][]{Magg2022, Kaur2022}.

\begin{comment}
\begin{figure}
    \centering
    \includegraphics[width=\columnwidth]{Figures/cdf_all_LyC.pdf}
    \caption{Cumulative distribution functions for various luminosity bands of interest in this paper. The curves are obtained by using the {\it mean} relation with respect to the halo mass. Different line-styles correspond to different redshifts and different colors represent different bands.}
    \label{fig:cdfs}
\end{figure}
\end{comment}

\subsubsection{Ionizing and Lyman Werner luminosities}
\label{sec:BPASS}

We use the Binary Population and Spectral Synthesis (BPASS) code to compute intrinsic ionizing and Lyman Wener UV luminosities \citep{Stanway2018, Byrne2022}.  %BPASS includes the effect of binary stars which increase the ionizing output for a fixed stellar age. The inputs for BPASS are the metalicity and star-formation history.  For the latter we assume that our sampled SFR is exponentially declining towards higher redshifts, as implied by Equation~\ref{eq:SFR}. 
BPASS provides a deterministic prediction for the UV luminosity, $L_{\rm UV}$, as a function of SFR, Z, and SFR history.  For the latter we assume that our sampled SFR is exponentially declining towards higher redshifts, as implied by Equation~\ref{eq:SFR}.   Therefore  $L_{\rm UV}$ is sampled assuming a log-normal conditional probability $p(\log L_{\rm UV} ~|~ \log {\rm SFR}, \log Z) = \mathcal{N}(\log L_{\rm UV} ~|~ \mu_{\rm UV, BPASS}({\rm SFR}, Z), \sigma_{\rm L})$ where $\mu_{\rm UV, BPASS}$ is the predicted luminosity from BPASS. We add an additional scatter of $\sigma_{\rm L} = 0.1$ dex around the mean to compensate for unaccounted sources of stochasticity, e.g. the mean IMF, alpha-element distribution, etc. \citep{Byrne2023}.  However, this level of scatter is negligible compared to the scatter of the bulk galaxy properties like SFR and stellar mass. We show the scaling relation of $L_{\rm UV}$ with $\rm SFR$  in the bottom panels of Fig.~\ref{fig:Lx_SFR} for the 11.2--13.6 eV Lyman-Werner band (left panel; in units of erg s$^{-1}$) and Lyman-limit (right panel; in units of erg s$^{-1}\AA^{-1}$ evaluated at 13.6 eV).
%etc. can create a scatter around. This is generally low and (REF) assume a value that is lower than $\sigma_{\rm UV, LW} = 0.1$. We take this value as out fiducial value for the bands in BPASS.

%The resulting luminosity distributions are shown in the lower panels of Fig.~\ref{fig:Lx_SFR}. We use two wavelength bands that are of interest at high-redshifts. In the lower left panel we show the soft-UV luminosity calculated in the Lyman-Werner band ($11.2$-$13.6$ eV), and in the lower right panel we show the specific luminosity evaluated at the Lyman-limit ($912\AA$).

%Luminosities at different wavelength bands show different scalings with SFR. Because of this, different sources of stochasticity will influence the emissivities in a different way. To facilitate the comparison in Sec.~\ref{sec:results} we show the cumulative contribution to the total photon budget as a function of halo mass for different wavelength bands, namely x-rays, LW photons and Lyman-limit photons in Figure~\ref{fig:cdfs}. X-ray luminosity shows a more top-heavy distribution compared to the UV luminosities. This is due to the steeper scaling with SFR in Eq.~\ref{eq:Lx-sfr}. Therefore, X-ray emissivity at high redshifts is going to be dominated by galaxies with larger SFRs, which become rarer at higher redshifts.

\subsection{Escape fractions}
\label{sec:escape_fraction}

Our final step in computing the emissivity is determining what fraction of the produced photons manage to escape the host galaxy into the IGM.  This is referred to as the escape fraction.  We use  different prescriptions for the escape fraction in our three bands of interest.  We describe each in turn below.

Both hydrodynamic simulations \citep[e.g.][]{Cen2015, Xu2016, Barrow2020, Yeh2022, Kostyuk2023} and direct observations of low redshift galaxies \citep[e.g.][]{Izotov2016, Grazian2017, Steidel2018, Pahl2023} show sizable stochasticity in the ionizing escape fraction, though there is no consensus on what is an appropriate distribution.
%\citep[though see the recent work of ][]{Kreilgaard2024}.  
Here we take two scenarios.  Our fiducial model assumes a log-normal distribution for the ionizing escape fraction with a width of 0.3 dex (c.f. \citealt{Mascia2023}), while we also show a bimodal distribution in which galaxies have an ionizing escape fraction of either 0 or 1 (resulting in maximum scatter).  In both cases we take the inference result of \citet{Nikolic2023}: $\bar{f}_{\rm esc}=0.053$ for the mean. We do not assume that the mean or scatter of the escape fraction depend on galaxy properties, as such dependencies are not yet well established for unbiased galaxy samples, especially at high redshifts. For example, assuming that scatter depends on mass could accelerate or decelerate reionization, by effectively shifting the population-averaged mean as a function of redshift.  Our framework can easily be extended to include putative dependencies on galaxy properties \citep[e.g.][]{Mascia2023}, as well as accommodating different functional distributions (e.g. \citealt{Kreilgaard2024}).  We defer such studies to future work.

For the X-ray escape fraction, we adopt the results of \citet{Das2017}, where they computed the X-ray opacities of simulated high-$z$ galaxies, finding that most photons with energies above 0.5 keV manage to escape.  Following that work, we assume an escape fraction of unity above 0.5 keV and zero below that value.  Similarly, we assume values of unity for the Lyman Werner escape fraction, given the typical low opacities of such photons through the host ISM  \citep[e.g.][]{Haiman2000, WolcottGreen2011}.

\section{Results: emissivities}
\label{sec:results}

Here we present our distributions for the emissivity in each of the three bands in turn.  We show the full distributions as a function of redshift, before quantifying the relative importance of each source of scatter.  For the latter, we compute the mean and standard deviations of the emissivity PDF when one source of scatter is removed (i.e. using only the corresponding mean relation with zero scatter), normalized to the values of the full distribution containing all sources of scatter: $\mu_{\epsilon}$/$\mu_{\epsilon}^{\rm full}$ and $\sigma_{\epsilon}$/$\sigma_{\epsilon}^{\rm full}$.  As mentioned above, we consider the following sources of scatter:
\begin{enumerate}[(i)]
\item spatial dependence of the mean CHMF on the large scale matter density
\item Poisson sample variance in halo number around the target mean CHMF
\item scatter around the SHMR
\item scatter around the SFMS
\item scatter around the FMR
\item scatter in the mapping of the intrinsic luminosity to SFR, $M_\ast$ and $Z$
\item scatter in the escape fraction
\end{enumerate}

We need to define a comoving volume over which to sum up the contributions of galaxies, in order to compute the emissivity PDFs.  Here we chose a fiducial scale of $R_{\rm nl} = 5$ cMpc.  This is roughly comparable to several relevant scales during the EoR and CD: (i) the typical HII bubble sizes during the early-middle stages of reionization \citep[e.g.][]{McQuinn2007, Lin2016}; (ii) the resolution of 21cm maps achievable after a 1000h observation with SKA1-low \citep[][]{Koopmans2015, Prelogovic2022}; (iii) the Lyman limit mean free path at $z\sim6$ \citep[e.g.][]{Becker2021}; and (iv) the field of view of {\it JWST} \citep[e.g.][]{Treu2022, Finkelsten2023, Bunker2023}.  Our emissivity PDFs are generated from $10000$ realizations of such volumes.  In Appendix~\ref{sec:appendix} we vary this scale and demonstrate that the estimated mean emissivities have converged to within a few percent.%, showing the emissivity distributions at $R_{\rm nl}$=1 cMpc (e.g. a typical cell size for large-scale EoR/CD simulations), as well as $R_{\rm nl}=10$ cMpc (e.g. a typical HII bubble size late in the EoR).

\begin{figure}
    \centering
    \includegraphics[width=1.0\columnwidth]{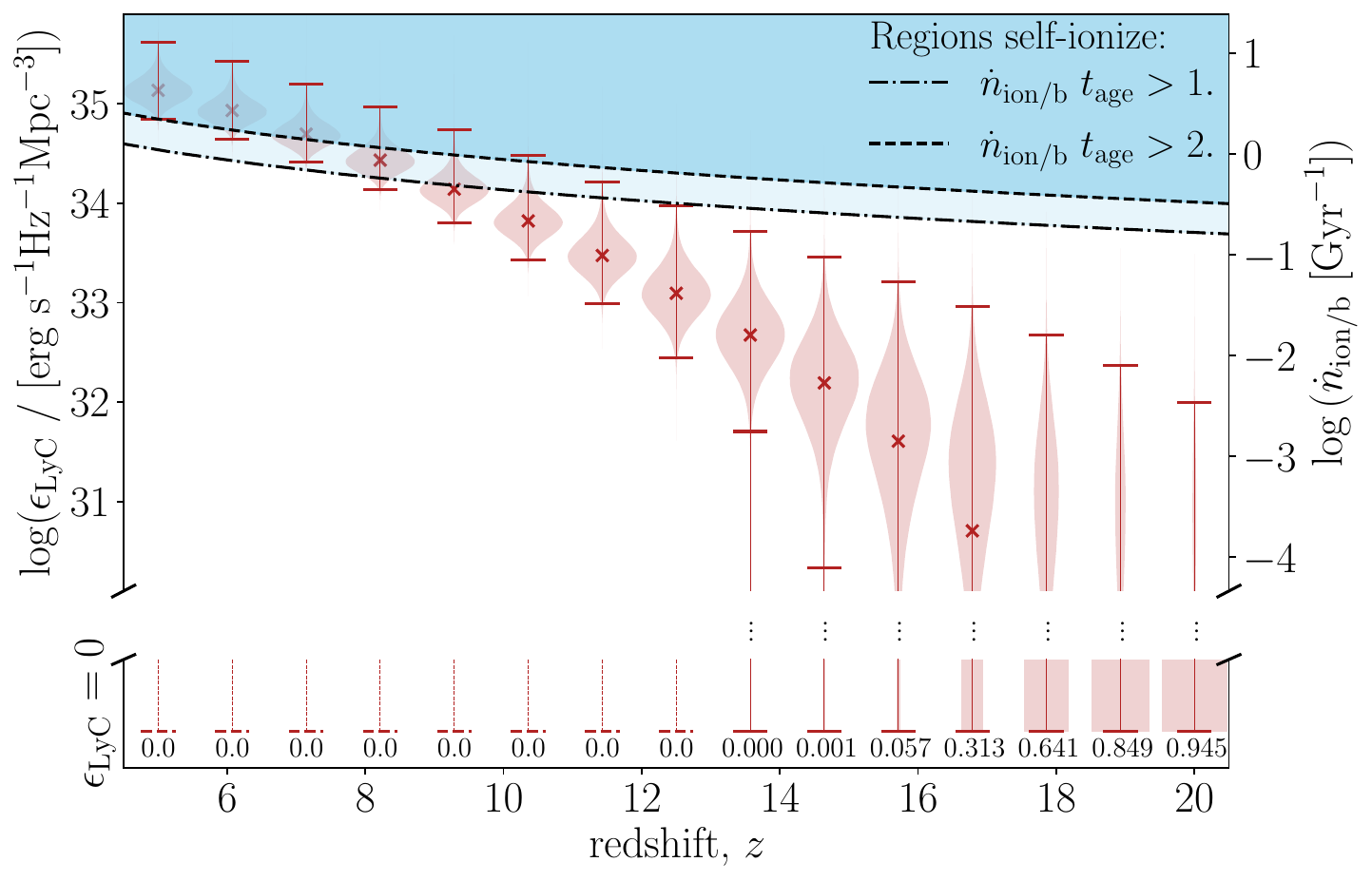}
    \caption{Distribution of Lyman limit  emissivities  for regions with a radius of 5 cMpc.  Violin plots correspond to the full emissivity PDFs, while the crosses and horizontal bars demarcate the mean and 99th percentiles, respectively.  The rectangle on the bottom with the matching number represents the fraction of 5 cMpc regions with zero emissivity.  On the left axis we show the specific emissivity at the Lyman limit, while on the right axis we show the corresponding number of ionizing photons (>13.6 eV) per baryon per Gyr.  The blue shaded region at the top demarcates the approximate criteria for a 5 cMpc to ionize: having an emissivity greater than one (dot dashed line) or two (dashed line) ionizing photons per baryon in the age of the Universe.  Assuming a threshold value of two ionizing photons per stellar baryon (e.g. \citealt{Bolton2007, Sobacchi2014}).) we see that roughly half of 5 cMpc regions can self-ionize by $z\sim$ 7, consistent with the latest estimates of the EoR history e.g. Qin et al. in prep}.
    \label{fig:violin_LyC}
\end{figure}

%-------------------------------------- Two column figure (place early!)

\subsection{Ionizing UV emissivity}
\label{sec:LyC_analysis}

In Figure~\ref{fig:violin_LyC} we show the distributions of the ionizing emissivity in our fiducial model, sampling all of the above-mentioned sources of stochasticity. 
 On the left axis we report the specific emissivity in units of erg s$^{-1}$ Hz$^{-1}$ evaluated at the Lyman limit, while on the right axis we show the total number of ionizing photons above the Lyman limit per baryon per Gyr. Red violins show the ionizing emissivity PDF, with crosses (horizontal bars) demarcating the mean (99\% C.L.) of the distributions. The fraction of our $R_{\rm nl}$ =  5 cMpc realizations that have a zero emissivity is denoted at the bottom of each violin.

\begin{figure*}[h]
    \centering
    \sidecaption
    \includegraphics[width=1.4\columnwidth]{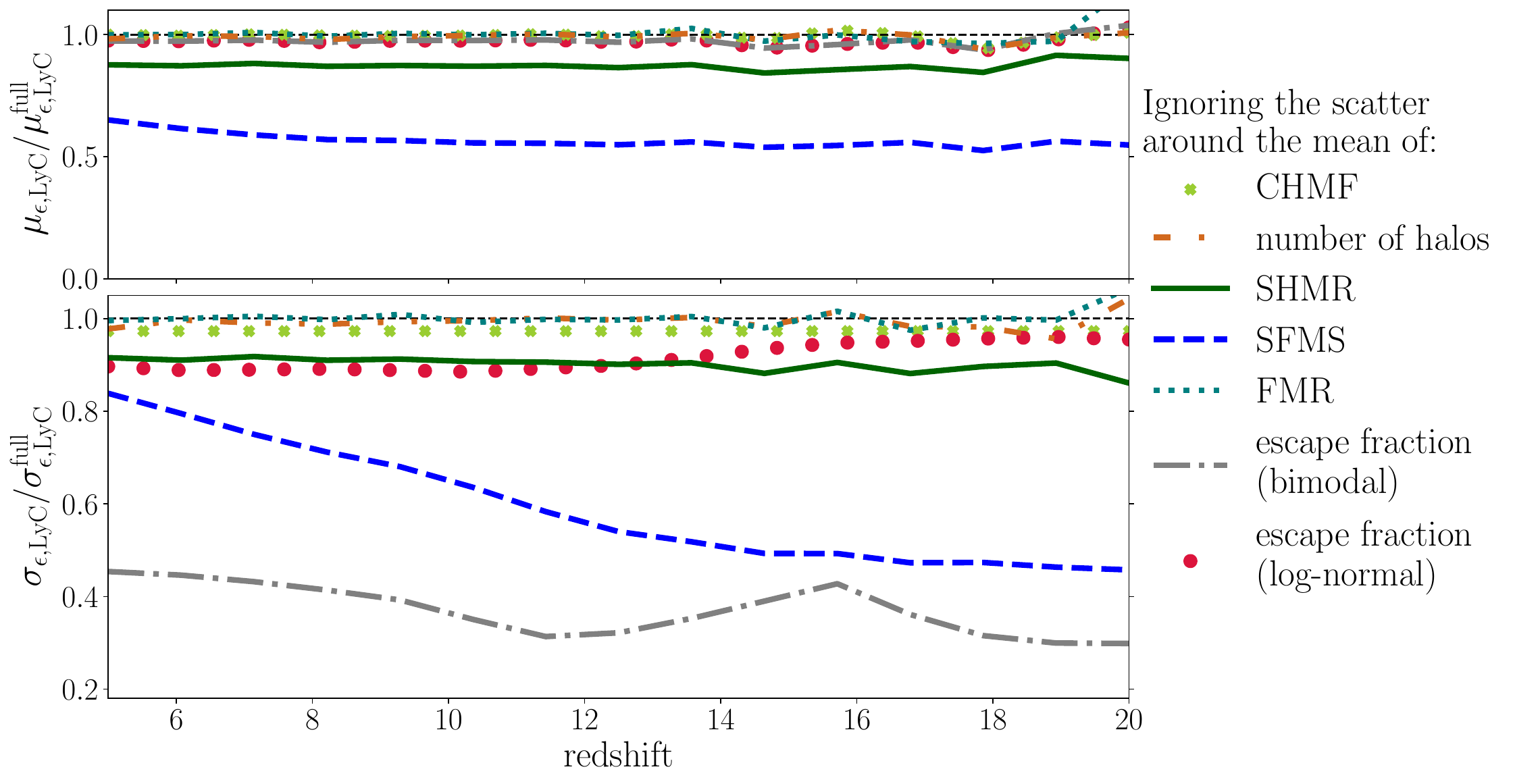}
    \caption{The fractional contribution of different sources of scatter (see legend) to the mean and standard deviation of the ionizing emissivity, computed over $R_{\rm nl}=5$ Mpc volumes. The top (bottom) panel shows the mean (standard deviation) when removing one source of stochasticity, normalized to the fiducial value that includes all scatter.}
    \label{fig:LyC-all}
\end{figure*}

 As galaxies become rarer towards higher redshifts, the mean ionizing emissivity decreases and the region to region scatter increases. The emissivity PDF becomes bimodal, with some regions having an emissivity of zero while those that have a non-zero emissivity show an approximately log-normal distribution. This is shown in Fig.~\ref{fig:LyC-all} with a rectangle at the bottom of the figure representing the probability that a region of $R_{\rm nl}=5$ cMpc has zero emissivity. At $z\gtrsim18$ the majority of $R_{\rm nl}$ =  5 cMpc volumes are expected not to have any galaxies that are actively emitting ionizing photons, in this fiducial scenario with a log-normal $p(f_{\rm esc})$.  If instead we assume a binomial $p(f_{\rm esc})$ distribution, the majority of $R_{\rm nl}$ =  5 cMpc have zero ionizing emissivity already by $z\gtrsim15$.
 
 %At redshifts $z>11$ there is an increasing probability that a region doesn't contain galaxies that emit ionizing galaxies that escape in the IGM. This is shown with a rectangle whose width is proportional to that probability. The exact value of the probability is given below each rectangle. Note that this probability is not the same as the probability that a region doesn't contain galaxies that \textit{produce} ionizing photons. We remind the reader that in our fiducial case, escape fraction is binomially distributed (see Section~\ref{sec:escape_fraction}), so that most of the galaxies will not contribute to the ionization of the IGM at a given time. The mean of the distribution is going down with redshift as is expected due to structure formation. On the other hand, the relative variance is increasing with redshift. This is caused by the decreasing number of halos with increasing redshift. It is for this reason that the effect of stochasticity becomes more important for higher redshifts.

We now quantify the main sources of scatter  driving the variance in Fig.~\ref{fig:violin_LyC}.  As discussed above, we do this by repeating our emissivity calculation but omitting one source of stochasticity (i.e. only using the corresponding mean relation with no scatter).  In Figure \ref{fig:LyC-all} we plot the corresponding mean ({\it top panel}) and standard deviation\footnote{For numerical stability, we calculate the standard deviation by fitting a log-normal to the non-zero distribution of emissivities.} ({\it bottom panel}), normalized to the corresponding values from the full calculation shown in Fig.~\ref{fig:violin_LyC}.  Note that since most of our sources of scatter are log-normal, assuming a mean relation instead of the full distribution would {\it underestimate} the mean of the emissivity shown in the top panel (see Appendix \ref{sec:appendix_B}).

%"We see the most important source of spatial variance in the ionizing emissivity is the spatial variation of the {\it mean} conditional halo mass function, driven by the large-scale matter field.   Ignoring this variance by assuming all 5 cMpc regions have on average the same halo mass function ({\it ?? curve}) would underestimate the mean and sigma by..... This is reassuring, since this scatter is already taken into account by all 3D semi-numerical and numerical simulations."

%First we note the effect large-scale density fluctuations have on the variance of the ionizing emissivity, i.e. variation of the conditional halo mass function. The variance changes only by a few percent. This is caused by the fact that the variance of the emissivity for a fixed density is larger than the change in the mean emissivity for a range of densities at this scale. Therefore any large-scale observable, such as the 21cm power spectrum, that depends on the density, will on small-scales have a non-negligible variance contribution coming from the stochasticity in the astrophysical processes. This will be seen as a shot-noise contribution at small scales.  

The most important source of scatter is the escape fraction, {\it if the escape fraction is binomial}.  In this scenario, assuming only the mean escape fraction for all galaxies would underestimate the standard deviation (std) by  60 -- 70\% throughout the EoR and CD ({\it gray dash-dotted curves}). However, the mean emissivity is unchanged (since the bimodal distribution has the same median and mean; see Appendix~\ref{sec:appendix_B}). 
If we instead assume that the escape fraction is log-normally distributed ({\it red circled curves}), not including scatter in this quantity only underestimates the mean and std of the emissivity by $\sim10 \%$.

Another important source of stochasticity is the burstiness of star formation.  Assuming all galaxies follow the mean SFMS without scatter ({\it blue dashed curves}) would underpredict the mean (std) of the ionizing emissivity by 40\% (15\%) at $z\sim5$.  This underprediction in the std rises to $\sim$ 50\% towards $z\sim20$, as the typical galaxies have smaller stellar masses and therefore a broader  $p({\rm SFR}~|~M_\ast)$ (c.f. the bottom panels in Fig. \ref{fig:SMHM-SFR}).  In other words, the increased "bursty" nature of star formation at higher redshifts (at which the emissivity is dominated by galaxies with smaller masses) drives a correspondingly larger spatial variance in the ionizing emissivity.

On the other hand, ignoring scatter around the SHMR results in an underprediction of the mean and std of the emissivity by only 10\%. Other sources of scatter have a negligible impact on the mean and variance of the ionizing emissivity.  In particular, we note that only
%Finally, the Poisson scatter in the halo number and the scatter around the FMR have a negligible contribution to the mean ionizing emissivity and its spatial variance.  Only 
$\sim5$\% of our realizations of 5 Mpc regions at $z=20$ contain fewer than 10 actively star forming galaxies.  Therefore, it is not surprising that Poisson scatter in the halo number is unimportant in determining emissivities.  We note that here we only consider galaxies above the atomic cooling threshold; had we considered an additional population of molecular-cooling galaxies, Poisson scatter would have been even less important since their expected mean number density is much larger.

\subsection{X-ray emissivity}
\label{sec:x-ray_analysis}
\begin{figure}[h]
    \centering
    
    \includegraphics[width=1.0\columnwidth]{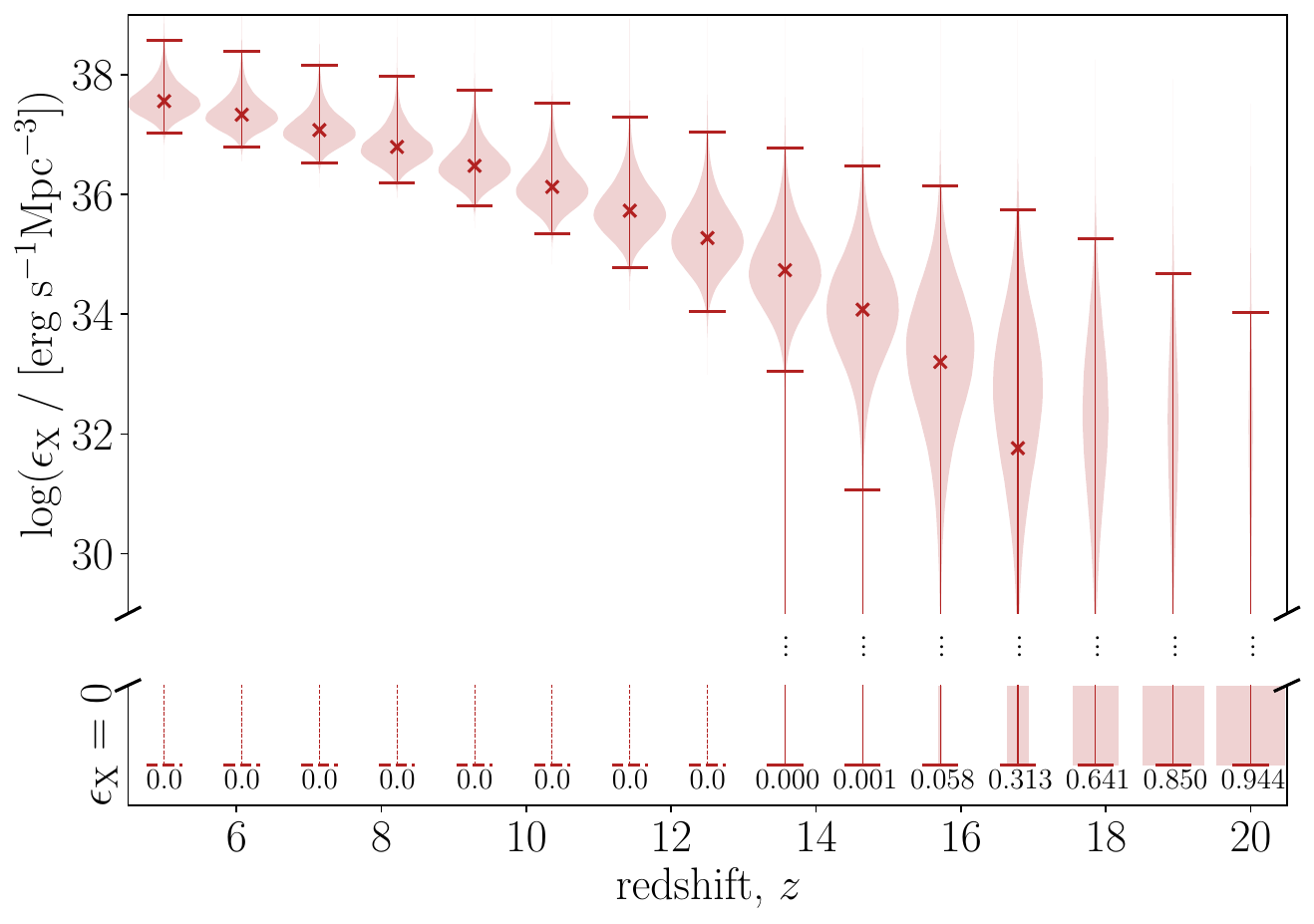}
    \caption{Like Fig. \ref{fig:violin_LyC}, but for soft band (0.5--2 keV) X-ray emissivity.}
    \label{fig:violin_x}
\end{figure}

\begin{figure*}
    \sidecaption
    \includegraphics[width=1.4\columnwidth]{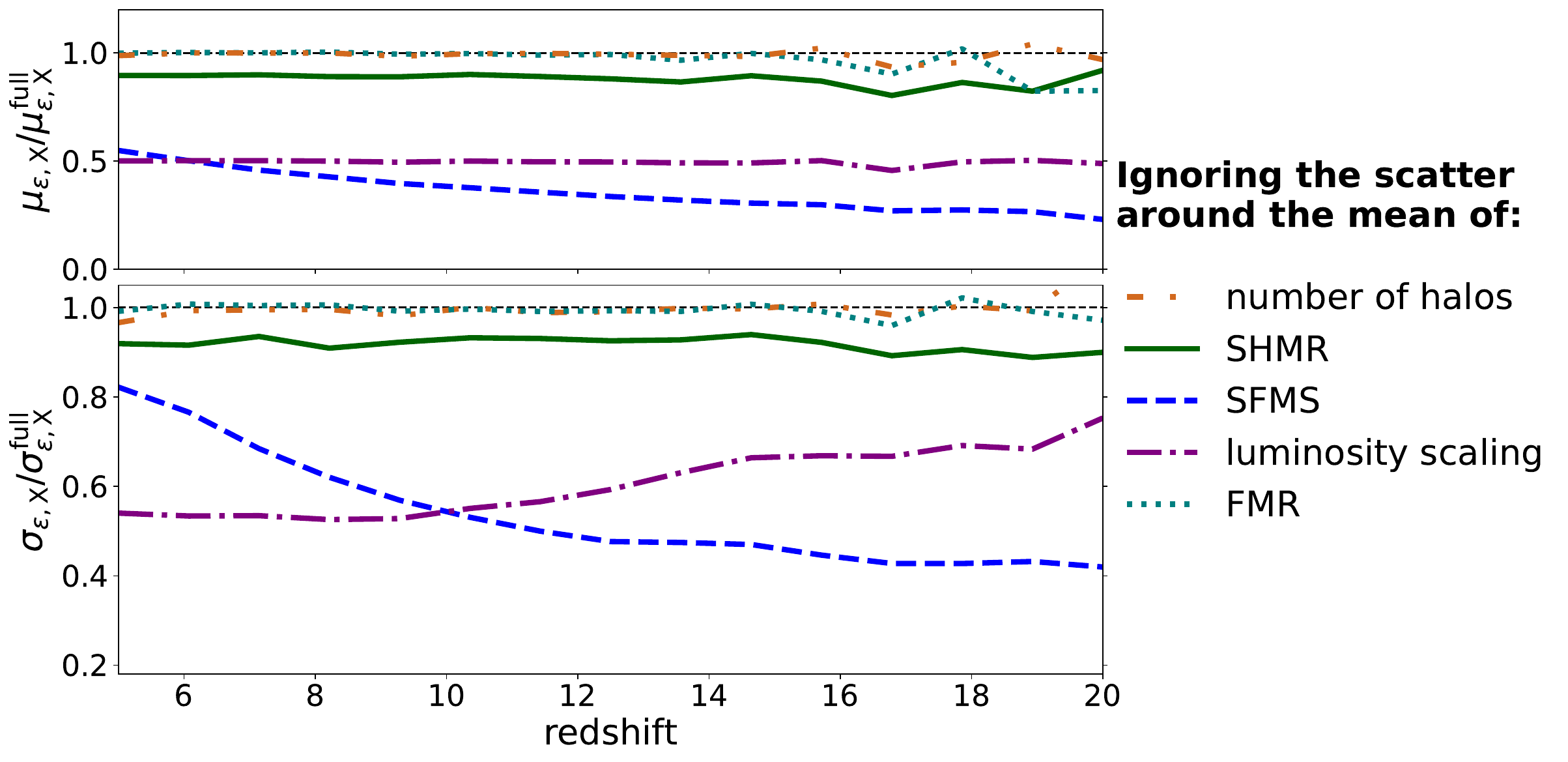}
    \caption{Like Fig. \ref{fig:LyC-all}, but for the  soft band (0.5--2 keV) X-ray emissivity.}
    \label{fig:X-all}
\end{figure*}
In Figure~\ref{fig:violin_x} we show the distributions of soft-band X-ray emissivities for our fiducial model, accounting for all sources of stochasticity. Red violins again represent PDFs averaged over comoving volumes with $R_{\rm nl}=$ 5 cMpc, while the rectangles represent probabilities that a region of size $R_{\rm nl}=5$cMpc has zero X-ray emissivity.  As expected the means and widths of the distributions decrease towards higher redshifts.  

We see that the X-ray emissivities have broader distributions compared with the ionizing emissivities in Fig. \ref{fig:violin_LyC}.  For example, at $z\sim10$ the region-to-region std of X-ray emissivities is $300\%$ of the mean, while for ionizing emissivities it is only $50\%$ of the mean.  This is primarily due to the fact that the HMXB LFs that source the X-ray emission in our model are fairly shallow \citep[see][]{Lehmer2021}.  Thus the galaxy-averaged X-ray luminosity is sensitive to sample variance as it can be determined by a small number of HMXBs.  This is evident by comparing the widths of the conditional $p(L ~|~ {\rm SFR}, M_\ast, Z)$ distributions for X-rays and ionizing photons in Fig. \ref{fig:Lx_SFR}.  The additional stochasticity in the ionizing emissivity due to the ionizing escape fraction (assuming it is log-normarly distributed) is sub-dominant compared with the wider X-ray intrisic luminosity distribution.
%regions devoid of X-ray sources start emerging at $z\gtrsim15$, whereas those devoid of ionizing sources emerge at $z\gtrsim11$.  Compared with ionizing photons, X-ray emission has more stochasticity in the intrinsic production (i.e. a wider $p(L ~|~ {\rm SFR}, M_\ast, Z)$ as seen in Fig. \ref{fig:Lx_SFR}), but less stochasticity in the ionizing escape fraction (which is taken to be unity for X-rays above 0.5 keV).  Comparing Figures \ref{fig:violin_x} and \ref{fig:violin_LyC}, we see that the stochasticity from the ionizing escape fraction is effectively larger than that from the HMXB luminosity function.

We isolate the relative importance of each source of scatter to the X-ray emissivity in Figure~\ref{fig:X-all}.  As in the previous section, we show  $\mu_i$/$\mu_{\rm full}$ and $\sigma_i$/$\sigma_{\rm full}$ in the upper and lower panels, respectively.  %Compared to Fig.~\ref{fig:LyC-all}, here these is no ionizing escape term, but there is an additional except for the exclusion of 'no f$_{\rm esc}$ scatter' curve which is not present given our assumption of x-ray escape fraction being unity. Instead we have the 'no Lx-SFR scatter' curve since x-ray scaling relation shows a great deal of scatter around the median.

The biggest impact on the mean and std comes from the scatter in the SFR$-M_{\ast}$ relation
(blue dashed curves).  Ignoring the scatter around the SFMS results in an underprediction of the mean (standard deviation) of the X-ray emissivity by 20\% at $z=5$ rising to a factor of $60\%$ at $z=20$.  As in the previous section, this is driven by the mass-dependence of scatter in SFMS.  The physical interpretation is the same: increased burstiness of star formation in small mass galaxies (that dominate at higher redshifts) boosts the variance of the X-ray emissivity.   The scatter around the SFMS is even more important for X-ray emissivity, compared with ionizing emissivity, due to the strong dependence of the intrinsic X-ray luminosity on the SFR (see Fig. \ref{fig:Lx_SFR} and associated discussion).

Another important source of scatter is the $L_{\rm X}-$SFR relation (violet dash-dotted curves). The relative difference in standard deviations is roughly $30\%$ at $z=20$, rising to $45\%$ at $z=5$. At $z=5$ it is more important than the scatter in SFR$-M_{\ast}$ relation. Note that complex physics of the formation of binary stars could induce an additional redshift dependence in this scatter, with different IMF's giving different populations of binary stars. This would go in the direction of increasing the importance of modeling $L_{\rm X}-$SFR scatter at earlier times.

Scatter in the SHMR has a $\sim10$\% effect, again without redshift dependence since we chose a constant width for $p(M_\ast ~|~ M_h)$.  Scatter in the other terms has a negligible impact on the X-ray emissivity.

%On the other hand we have metaliticty scatter which doesn't alter the variance significantly. This is because of the low level of scatter predicted, $\sigma_{Z}$, but also because X-ray luminosity is only weakly dependent on metalicity, as shown in Fig.~\ref{fig:Lx_SFR}. The same is true for Poisson sampling of the number of halos which doesn't change the distribution significantly.

\subsection{Lyman Werner emissivity}
\begin{figure}
    \centering
    \includegraphics[width=1.0\columnwidth]{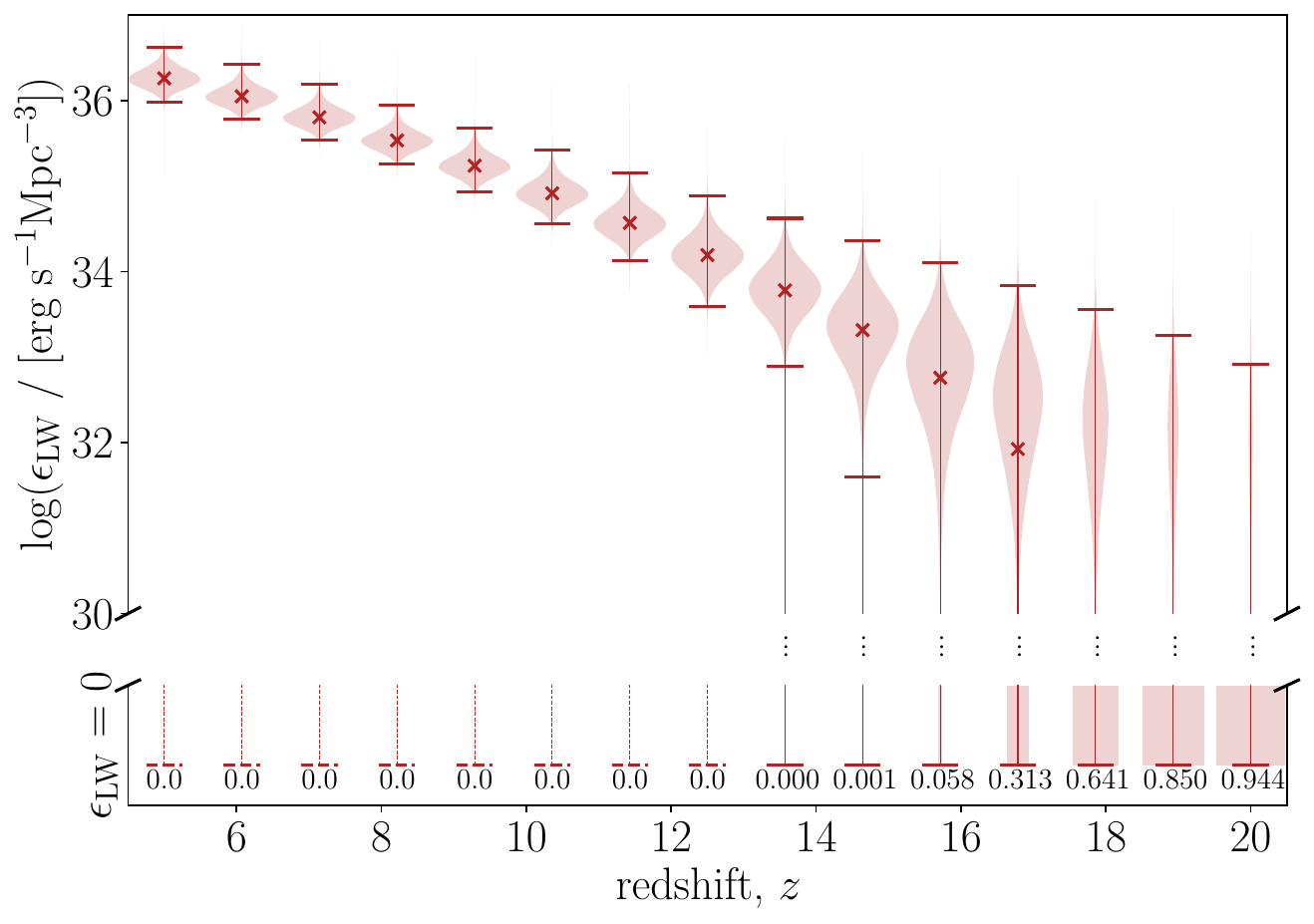}
    \caption{Like Fig. \ref{fig:violin_LyC}, but for the Lyman-Werner (11.2-13.6 eV) emissivity.}
    \label{fig:LW_violin}
\end{figure}

\begin{figure*}
    %\centering
    \sidecaption
    \includegraphics[width=1.4\columnwidth]{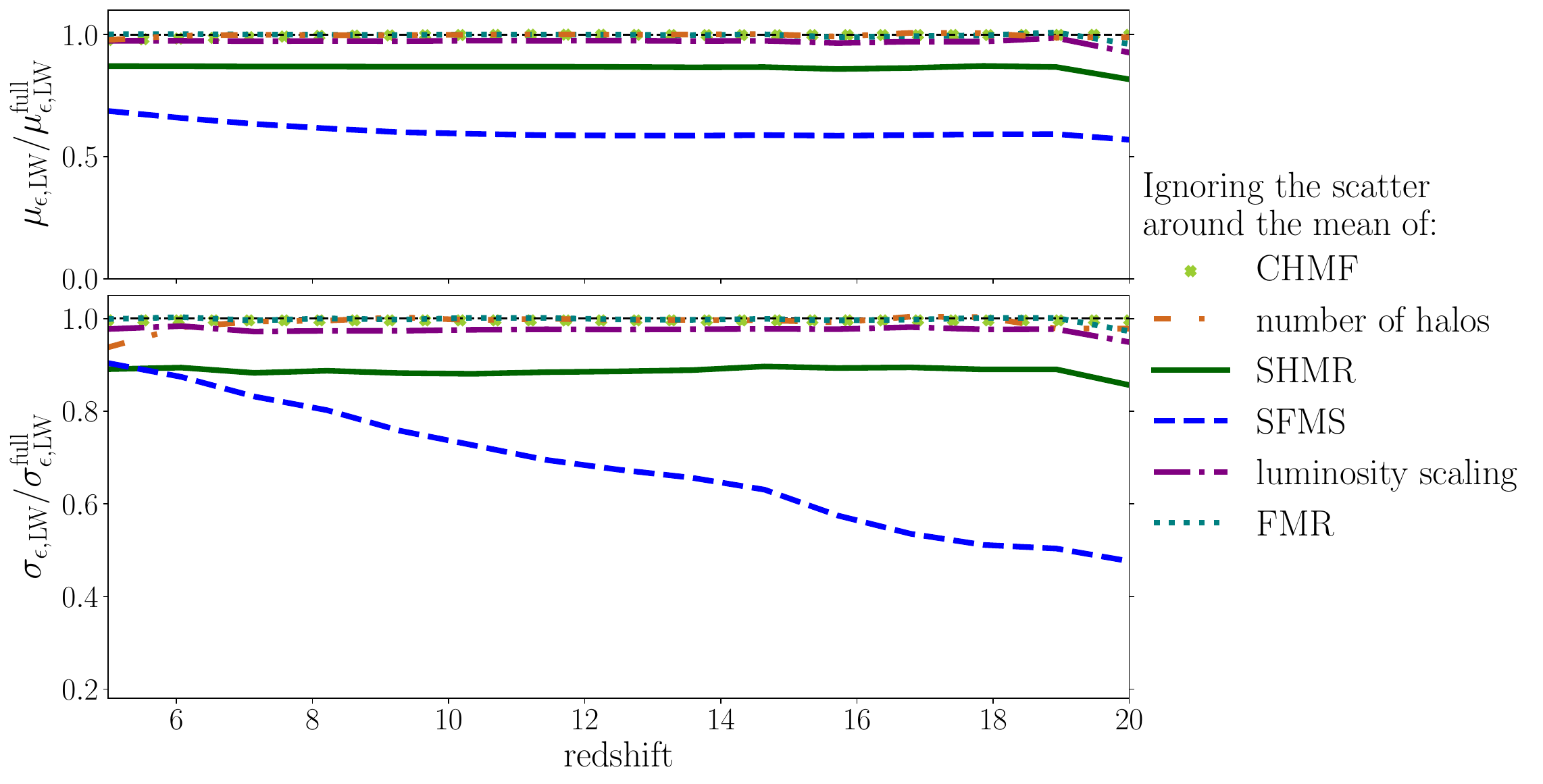}
    \caption{Like Fig. \ref{fig:LyC-all}, but for the  Lyman Werner (11.2-13.6 eV) emissivity.}
    \label{fig:LW-all}
\end{figure*}

Soft UV photons are important during the cosmic dawn as they regulate the abundances of H$_2$ (which provides an important cooling channel for the first galaxies) and the excited spin state of HI (which determines the cosmic 21cm signal).  For concreteness, here we evaluate the emissivity in the Lyman-Werner band ($11.2$-$13.6$eV) noting that our conclusions would remain the same regardless of the specific soft UV range of interest.
 
In Fig. \ref{fig:LW_violin} we show the distribution of LW emissivities in 5 cMpc regions.  We see that LW emissivities are more uniform (i.e. with narrower PDFs) than both ionizing or X-ray emissivities from the previous subsections.  This is to be expected, as the latter bands are sensitive to stochasticity in the ionizing escape fraction and HMXB LFs, neither of which contribute to the LW emissivity. 

In Fig. \ref{fig:LW-all} we show the fractional contribution of different sources of scatter to the mean and std of LW emissivity.
Again, the SFMS (blue dashed lines) is the most important contributing source to the variance of emissivity, but less so compared to X-rays (as could be expected from Fig. \ref{fig:Lx_SFR}). At $z\sim20$ the burstiness of SFR contributes at a $\sim 50$\% level to the std of the distribution, but this drops to $\sim 10$\% at $z\sim 5$. %Physical interpretation is again similar, with decreasing redshift typical halo has a larger mass, and a lower scatter in the SFMS.  

Also important is the scatter in the SHMR (green full curves) which contributes at a $\sim 10$\% level to the mean and std for all redshifts.  Other sources of scatter are negligible.

%\section{Discussion}

%\subsection{Some assumptions}

%Test some assumptions like the radius of the region.

%Comment on the difference with respect to flaring out. Comment on the \citet{Hassan2021} paper and comparisons to other works.
%
%                               
%-----------------------------------------------------------------
\section{Results: EoR history}
\label{sec:EoR}

\begin{figure*}[h]
    \centering
    \includegraphics[width=\textwidth]{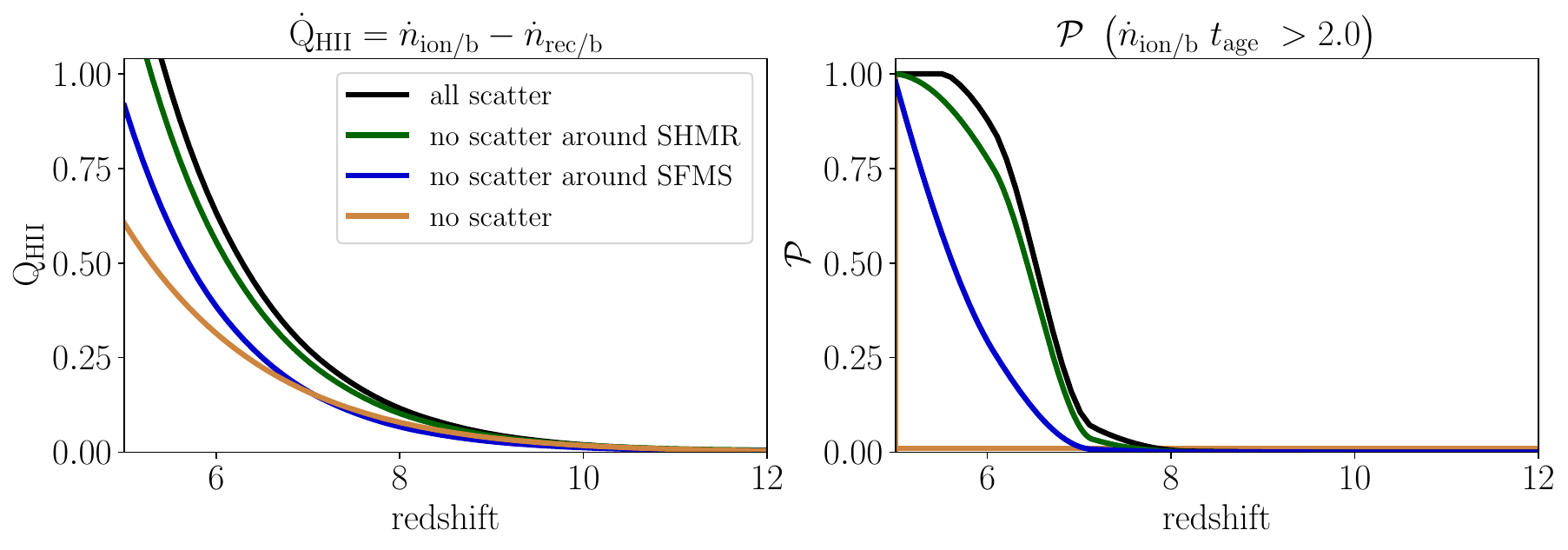}
    \caption{ The relative contribution of galaxy stochasticity to the EoR history.  In the left panel we show the common approximation of an EoR history calculated assuming a constant clumping factor (c.f. Eq.~\ref{eq:EoR_eq}), while in the right we show the fraction of 5 cMpc regions that exceed the threshold of two ionizing photons per baryon per age of the Universe (c.f. Fig. \ref{fig:violin_LyC}). Black curves correspond to our fiducial model, including all sources of scatter.  Green / blue  curves ignore scatter around the SHMR / SFMS, while the orange curves ignore all sources of scatter.  We see that not accounting for the burstiness of star formation (i.e. scatter around the SFMS) can result in EoR histories that are delayed by $\Delta z\sim$ 0.5--1.  Not accounting for any scatter and assuming only mean galaxy properties delays the completion of the EoR by $\Delta z \sim$ 2.}
    \label{fig:EoR_hists}
\end{figure*}

The ionizing emissivities shown in the previous section can be used to estimate the redshift evolution of the volume filling factor of ionized regions, $Q_{\rm HII}(z)$  -- the EoR history.  Even though it is an average quantity, computing the EoR history accurately requires accounting for the spatial and temporal co-evolution of sources and sinks of ionizing photons, and is therefore best done numerically.  However popular analytic approximations exist and can provide insight into the relative impact of scatter in galaxy properties.

Here we compute two proxies for the EoR history.  The first is the most common approximation in the literature \citep[e.g.][]{Madau1999}, obtained by:
\begin{equation}
    \frac{\textrm{d}Q_{\rm HII}}{\textrm{d}t} =\dot{n}_{\rm ion/b} - \alpha_{\rm A} \ C \ \langle n_{H} \rangle \ Q_{\rm HII}.
\label{eq:EoR_eq}
\end{equation}
Here $\dot{n}_{\rm ion/b}$ is the ionizing emissivity per baryon predicted by our model\footnote{We make the standard assumption that helium is singly ionized by stellar sources together with hydrogen, due to their comparable ionization thresholds.}, $\alpha_{\rm A}$ is the case-A recombination coefficient, $\langle n_{H} \rangle$ is the mean hydrogen density, and $C\equiv \langle n_H^2 \rangle / \langle n_{H} \rangle^2$ is the so-called "clumping factor" computed only over the ionized (not self-shielded) gas.  By assuming a constant clumping factor, this equation ignores the correlation between sources and sinks of ionizing photons.\footnote{In reality, most recombinations will come from the earliest patches of the IGM to ionize, which are those with the highest densities of galaxies.  As a result, the growth of HII regions surrounding the highest galaxy densities begins to stall as reionization progresses, with an increasing fraction of ionizing photons required to balance recombinations.  This process naturally results in a "soft landing", with the recombinations starting to balance ionizations in the late EoR stages, smoothly transitioning to the post-EoR regime (e.g. \citealt{Sobacchi2014}).}  Estimates of the EoR history obtained with eq. (\ref{eq:EoR_eq}) underpredict the duration of the EoR by $\Delta z \sim$ 1--2, with the error increasing towards the end stages (see, e.g. Figure 6 in \citealt{Sobacchi2014}).  Here we take $C=2$, noting that we are only interested in the {\it relative} impact of galaxy stochasticity on the EoR history.  

We show the resulting estimates in the left panel of Fig.~\ref{fig:EoR_hists}.  The black curve corresponds to our fiducial model, in which we account for all of the aforementioned sources of scatter.  The green (blue) curve is computed ignoring scatter around the mean SHMR (SFMS).  The orange curve does not account for any scatter, taking only the mean values for each relation.  We see that scatter in the SHMR only delays the EoR history by $\Delta z \sim$ 0.1.   Ignoring scatter around the SFMS has a bigger impact, delaying the EoR history by $\Delta z \sim$ 0.5 -- 1.
Ignoring scatter in all galaxy properties underestimates the duration of the EoR and delays the end stages by up to $\Delta z \sim$ 2.

Our second proxy for the EoR history is obtained directly from Fig.~\ref{fig:violin_LyC}. Specifically, we compute the fraction of 5 cMpc regions whose emissivities are larger than two ionizing photons per baryon per age of the Universe at that redshift, i.e. $\dot{n}_{\rm ion/b}  t_{age}> 2$ (shown by the dashed line in Fig~\ref{fig:violin_LyC}; c.f. \citealt{Bolton2007, Sobacchi2014}).  This approximation of the EoR history assumes that each 5 cMpc region of the Universe is instantaneously ionized when this criterion is reached, and that each such region is independent.  However, it does correctly compute the spatial variation in the emissivity, allowing  us to account for local recombinations by increasing the required ionizing photon threshold.   Again, we stress that here we are only interested in the relative impact of galaxy stochasticity on the EoR history.

We show the resulting estimate in the right panel of Fig.~\ref{fig:EoR_hists}, for the same models as shown in the left panel.  The qualitative evolution of this quantity is different from the one in the left panel.  By its definition, taking only mean values (orange curve) would result in a step function at the end of the EoR.
 Importantly however, the relative impact of ignoring scatter in the SFMS and SHMR is the similar in both panels.

Regardless of the proxy used in estimating the EoR history, neglecting scatter around the SFMS results in a delayed EoR history by $\Delta z \sim$ 0.5--1.  Neglecting {\it all} galaxy-to-galaxy scatter by assuming mean values (c.f. Eq.~\ref{eq:simple-emissivity}) results in an overextended EoR history and delays its completion by $\Delta z \sim$ 1--2.  

Our results suggest that inferring galaxy properties from EoR history data without accounting for stochasticity could bias recovery towards brighter galaxies or higher escape fractions.  We will investigate this further in future work, using 3D simulations that can more accurately capture the evolution of photon sinks and thus better predict the EoR history.

\section{Results: UV luminosity functions}
\label{sec:LFs}

\begin{figure*}
    \centering
    \includegraphics[width=\linewidth]{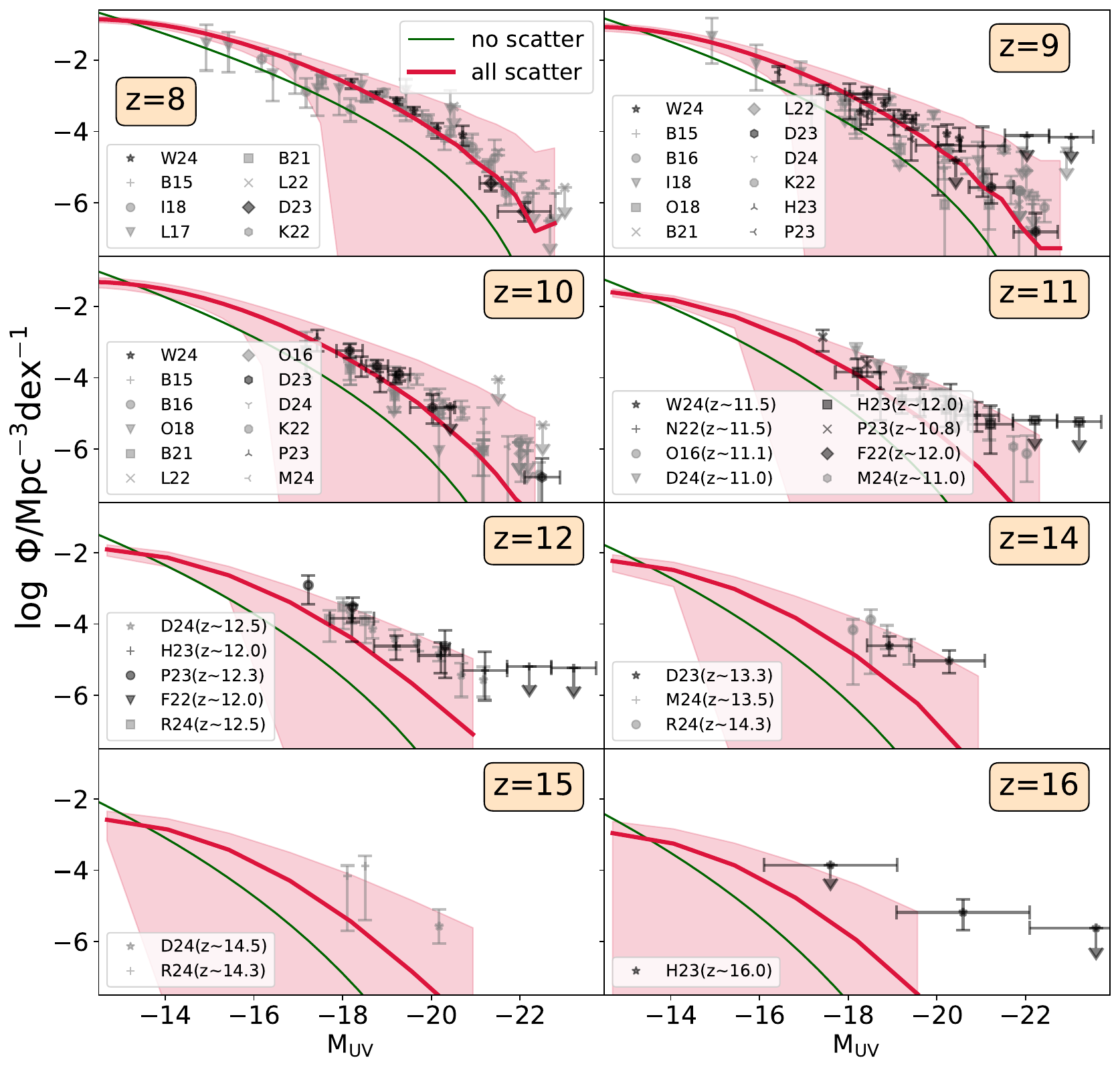}
    \caption{High redshift UV luminosity functions.  Our fiducial model including all of the aforementioned sources of scatter is shown with the red solid lines ({\it mean values}) and surrounding shaded regions ({\it 68\% C.L.s}).  The solid green curves correspond to UV LFs calculated using only mean relations without any scatter.  Also shown in each panel are various observational estimates from {\it HST} and {\it JWST} (see text for details).}
    \label{fig:uvlfs}
\end{figure*}

The framework developed in Section ~\ref{sec:emissivities} also allows us to compute the corresponding galaxy UV LFs.  Specifically, for each galaxy realization, we measure its rest frame magnitude using the luminosity from $1450\AA$ to $1550\AA$ directly from BPASS (see Section~\ref{sec:BPASS}).  For simplicity, we do not account for nebular emission, nor dust attenuation \citep[e.g.][]{Ferrara2023}.  We will include these in future work focused on interpreting UV LFs.

In Fig. \ref{fig:uvlfs} we plot the mean UV LF at each redshift (red line) along with the 68\% C.L. (red shaded region). In green, we show the UV LFs calculated assuming only mean relations without any scatter.
Also shown are various observational estimates from {\it HST} (\citealp{Bouwens2015} (B15); \citealp{Bouwens2016} (B16); \citealp{Bouwens2021} (B21); \citealp{Livemore2017} (L17); \citealp{Ishigaki2018} (I18); \citealp{Oesch2016} (O16); \citealp{Oesch2018} (O18); \citealp{Leethochawalit2022} (L22); \citealp{Kauffmann2022} (K22)) and {\it JWST} (\citealp{Naidu2022} (N22); \citealp{Finkelsten2023} (F22); \citealp{Donnan2023} (D23); \citealp{Donnan2024} (D24); \citealp{PerezGonzalez2023} (P23); \citealp{Robertson2023} (R24); \citealp{Harikane2024} (H23); \citealp{Mcleod2024} (M24); \citealp{Willott2024} (W24)).

% \citep{Bouwens2015, Bouwens2016, Bouwens2021, Livemore2017, Ishigaki2018, OEsch2016, Oesch2018, Leethochawalit2022, Kauffmann2022} and JWST \citep{Naidu2022, Finkelstein2023, Donnan2023, Donnan2024, PerezGonzalez2023, Robertson2023, Harikane2024, Mcleod2024,  Willott2024}.

Comparing the green and red curves, we see that including scatter shifts the mean to brighter magnitudes and flattens the UV LFs.  This is a well-known effect of upscattering some fraction of the more abundant faint galaxies to brighter magnitudes.  For our fiducial model, the shift is roughly 1-2 magnitudes at $M_{\rm UV} \sim$ -18, consistent with other estimates of the impact of stochasticity on UV LFs \citep[e.g.][]{Mason2023, Shen2023, Gelli2024}.

Comparing the red curve to the observational estimates, we see that our fiducial model is consistent with UV LFs at $z\lesssim10$.   However, the mean underpredicts the recent estimates of UV LFs at $z \gtrsim 11$ from broad-band {\it JWST} photometry.  Although the observational data points are mostly within the 68\% C.L. of our model, they are {\it systematically} higher.  Assuming there is no observational bias and that there are no correlations between the magnitude bins, this systematic underprediction would imply that our fiducial model is strongly disfavored by the data at $z \gtrsim 11$.  
This is qualitatively consistent with previous conclusions from the literature that larger than expected levels of scatter would be required to explain {\it JWST} results, provided the $z>10$ photometric estimates are accurate (e.g. \citealt{Mirocha2022, Mason2023, Shen2023, Pallottini2023, Gelli2024}).
Alternately, a correlation between observational estimates in different magnitude bins \citep[for example through cosmic variance; e.g. ][]{Willott2024} could alleviate this apparent tension.

\begin{figure}
    \centering
    \includegraphics[width=1.0\linewidth]{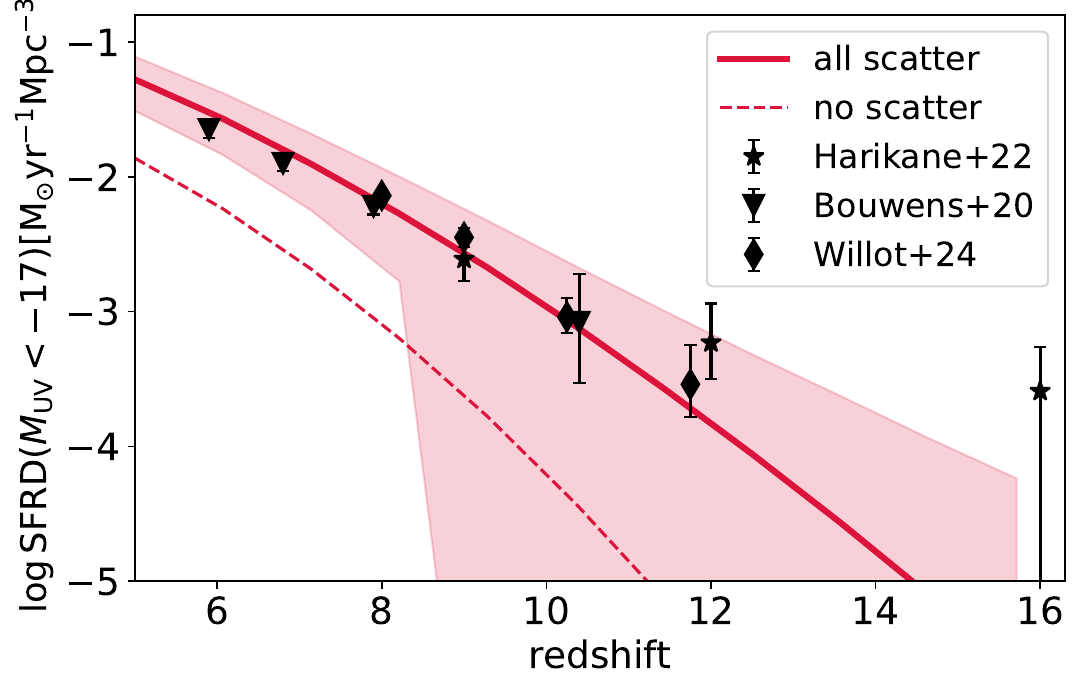}
    \caption{Star-formation rate density from bright galaxies ($M_{\rm UV} < -17$) as a function of density. Our fiducial model including all of the aforementioned sources of scatter is shown with the solid line and surrounding shaded regions ($68\%$ C.L.). The dashed line corresponds to the SFRD calculated using only mean relations without any scatter. Also shown are high-redshift estimates from \citet{Bouwens2020,Harikane2023, Willott2024}.}
    \label{fig:SFRD}
\end{figure}

We also calculate the star-formation density evolution from galaxies down to $M_{\rm UV} < -17$.  In order to provide a more like-to-like comparison with observations, we do not use the sampled SFRs directly but instead convert from the UV luminosity via a constant conversion factor: SFR$ (M_{\odot}\textrm{yr}^{-1}) = K_{\rm UV} L_{\rm UV}(\textrm{erg s}^{-1} \textrm{Hz}^{-1})$, where $K_{\rm UV}$ is the conversion factor which depends on the IMF and star-formation history. We take $K_{\rm UV}=1.15 \times 10^{-28} M_{\odot}\textrm{yr}^{-1}/\textrm{ergs}^{-1}\textrm{Hz}^{-1}$ \citep{Sun2016}, consistent with other works.  The result is shown in Fig.~\ref{fig:SFRD}. The solid curve represents our fiducial model which includes all sources of stochasticity, while the dashed curve represents the model where only the mean relations are considered. We see that our model reproduces the data very well at $z\lesssim12$ \citep{Bouwens2020}, though is somewhat lower than some estimates at higher redshifts.  As this statistic is fundamentally an integral over the UV LFs, we reach the same qualitative conclusions.  Quantitatively, the discrepancy with the data seems less than for the UV LFs.  Due to the steepness of the UV LFs, the integrated SFRD is dominated by the faint end limit used to compute it, while the $z>12$ JWST observations are more discrepant on the bright end.

The empirical framework we developed here is very flexible, and allows us to explicitly define the mean and scatter in every fundamental relation that leads to the 1500 \AA\ UV magnitude.  In future work we will use our model combined with physically-motivated priors to {\it infer} these conditional distributions from {\it JWST} UV LFs and other observational data.

\section{Conclusions}
\label{sec:conclusions}

In this work we quantify how does the galaxy-to-galaxy scatter in their properties impact estimates of their emissivities and related observables.  We use a semi-empirical model that explicitly defines scatter around well-studied mean relations: (i) the conditional halo mass function (CHMF); (ii) the stellar-to-halo mass relation (SHMR); (iii) galaxy star formation main sequence (SFMS); (iv) fundamental metallicity relation (FMR); (v) conditional intrinsic luminosity; and (vi) photon escape fraction.  We compute the corresponding multi-frequency (ionizing UV, X-rays, LW) emissivities, EoR histories, and UV LFs, quantifying the relative importance of the above sources of scatter.

We find that the burstyness of star formation (i.e. scatter around the mean SFMS) is important for all emissivities.  Because we assume burstiness increases towards smaller mass galaxies, the scatter around the SFMS  becomes increasingly important at higher redshifts.
Neglecting this source of stochasticity could underpredict the mean and std of emissivities by factors of up to few-10 during the EoR and CD.  Stochasticity in the ionizing escape fraction can dominate the spatial scatter in the ionizing emissivity {\it if} its distribution is binomial.  If instead the escape fraction is log-normally distributed, its contribution to the total emissivity scatter is only of order $\sim$ 10\%.  For the X-ray emissivity, one must account for scatter in the intrinsic luminosity, which in our fiducial model is driven by high mass X-ray binary luminosity functions.

We find that neglecting stochasticity overestimates the duration of reionization, delaying its completion by $\Delta z \sim$ 1--2.  Neglecting only scatter around the mean SFMS results in a delay of the EoR history by $\Delta z\sim$ 0.5 -- 1.   This suggests that inferring galaxy properties from EoR history data without accounting for stochasticity could bias recovery towards brighter galaxies or higher escape fractions. 

We recover the well-known effect of stochasticity flattening the UV LFs.  In our fiducial model, this results in a shift of 1--2 UV magnitudes at $M_{\rm UV} \sim$ -18. Our UV LFs are consistent with observational data at $z\leq10$ but consistently under-predict recent estimates at at higher redshifts.   This is qualitatively in line with other studies, and implies that larger scatter is required in order for it to be the sole explanation for photometric estimates at $z>10$.

We conclude that models of the EoR and CD should at least account for scatter around the SFMS.  Simulating the X-ray background during these epochs (for example when computing the 21cm signal) additionally requires accounting for scatter in the intrinsic X-ray luminosities of galaxies.  

The semi-empirical framework we use here is flexible and transparent.  It can easily be extended to accommodate additional observables, different functional distributions, and/or dependencies on additional galaxy properties.

\section{Data availability}
The code related to the work is publicly available at \faGithub \href{https://github.com/IvanNikolic21/Stochasticity_sampler}{IvanNikolic21/Stochasticity Sampler}.

\begin{acknowledgements}
We thank the anonymous referee for their insightful comments. We thank Yuxiang Qin for the help with collecting UV LF observations. We gratefully acknowledge computational resources of the Center for High Performance Computing (CHPC) at SNS.  A.M. acknowledges support from the Italian Ministry of Universities and Research (MUR) through the PRIN project "Optimal inference from radio images of the epoch of reionization" as well as the PNRR project "Centro Nazionale di Ricerca in High Performance Computing, Big Data e Quantum Computing".
\end{acknowledgements}

\bibliographystyle{aa}

\bibliography{stoc}

\appendix

\section{Convergence of the mean emissivity with scale}

\label{sec:appendix}
\begin{figure}[h]
    \centering
    \includegraphics[width=\linewidth]{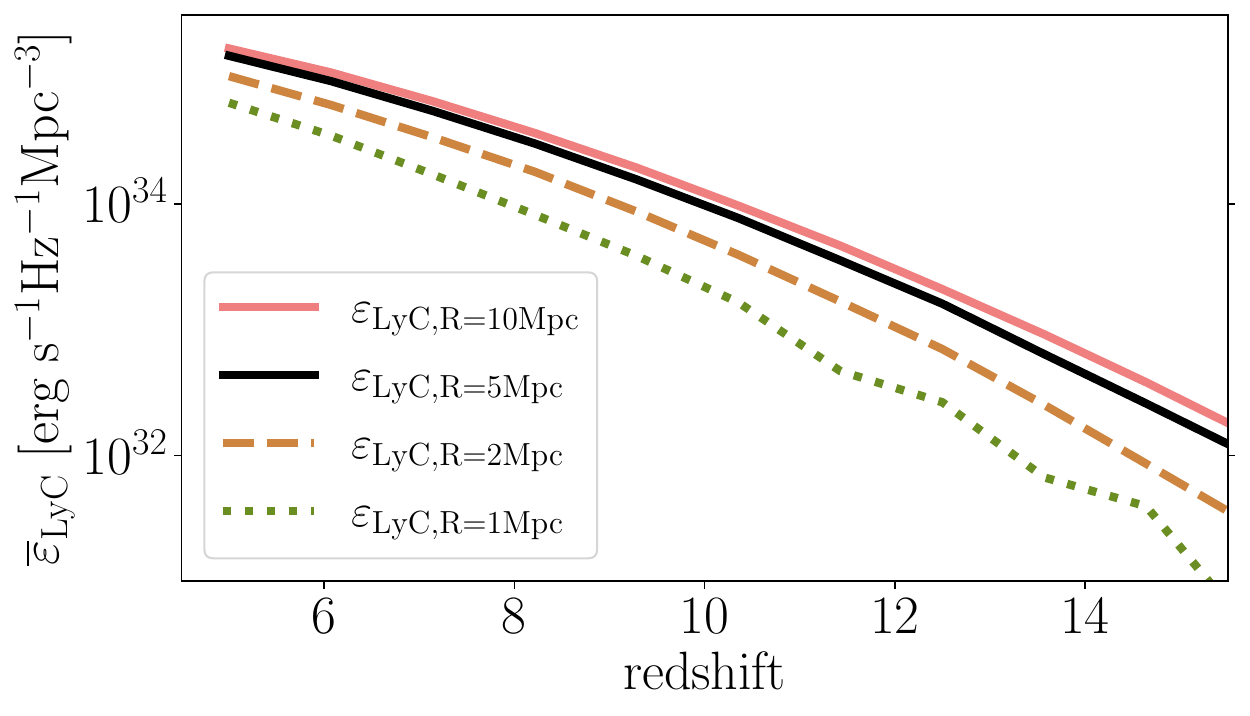}
    \caption{The mean of the ionizing emissivity computed over regions of varying scales, $R_{\rm nl}$ = 1, 2, 5, 10 cMpc. Our fiducial choice of 5 cMpc has converged in the mean to within a few percent. The analytical excursion set model misses massive halos when conditioned smaller scales, resulting in an underprediction of the mean for $R_{\rm nl}$ = 1 cMpc by factors of few -- 10. }
    
    \label{fig:varying_radii}
\end{figure}

Here we confirm that our fiducial choice of $R_{\rm nl}$ = 
 5 cMpc when computing emissivity distributions converges to the correct mean.  In principle, the averaging over the 
 scale-dependent overdensity distribution, $p_z(\delta_0 | R_{\rm nl})$, in the top row of Eq. \ref{eq:full_emissivity} should self-consistently ensure that the correct mean is recovered, regardless of the choice of scale.  In practice however, conditional halo mass functions underpredict the numbers of relatively massive halos whose Lagrangian volumes are close to the conditioning scale.

In Figure~\ref{fig:varying_radii} we illustrate how the mean emissivity changes with scale.  We use the ionizing emissivity to illustrate the trend; however, the result is the same for the other two bands of interest. We chose three additional scales: 1, 2, 10 Mpc.  The first two choices roughly span the cell sizes used in semi-numerical \citep[e.g.][]{Mesinger2011,Munoz2022, Schaeffer2023}, or low-resolution \citep[e.g.][]{Dixon2016, Meriot2024} radiative transfer simulations of the EoR/CD, while the latter roughly corresponds to the HII bubble size late in the EoR.

Comparing the red and black curves, we see that the mean for our fiducial choice of 5 cMpc has converged to within a few percent.  We are thus reassured that our fiducial choice can be used to predict global quantities like the EoR history and UV LFs.

As the scale is further reduced, we see that the mean emissivity can be significantly underestimated when using conditional excursion set formalism.  For example, the mean using $R_{\rm nl} = $ 1 cMpc is underestimated by factors of few - ten. 
This serves as a caution against computing halo fields {\it only} at the cell level for low-resolution EoR/CD simulations (e.g. Appendix A in \citealt{Davies2022b}; \citealt{Ries2022}).  Instead, $N$-body (e.g. \citealt{Dixon2016, Schaeffer2023, Meriot2024}), excursion-set that accounts for larger scales \citep[e.g.][]{Furlanetto2004, Mesinger2011, Cen2015}, or a mixture of the two \citep[e.g.][Davies et al. in prep]{McQuinn2007} should be used. 

\begin{comment}
The mean evolutions caution against using cell averaged quantities to predict the mean emissivity or EoR history analytically (since the product of the means does not equal the mean of the products; see Appendix \ref{sec:appendix_B}).

the change in the mean and variance with the change of the radius of the region. This is shown in Figure~\ref{fig:varying_radii}. We use R$=5$Mpc again as our fiducial choice and report fractional change with respect to emissivity at that size. We use x-rays to demonstrate the effect this has no the mean and variance, but the result is the same for the ionizing and LW bands.

Fig.~\ref{fig:varying_radii} indicates that the smaller the region is, the smaller is the mean and bigger is the variance. The mean changes mostly because we are not sampling higher-density modes. This is inherent in the the formalism of conditional halo mass functions. On the other hand, the variance increases with decreasing radius. The smaller is the region, there are less halos inside over which the emissivities is calculated. Therefore there is a strong variation from region to region. When we get to scales of R=$1$Mpc, which is a typical resolution of semi-numerical simulations, the standard deviation increases by a factor of $20$ for $z=5$ and that number increases to $100$ for $z=20$.
\end{comment}

% \renewcommand*\thesection{\Alph{section}}

\section{Shift in the mean emissivity for correlated log-normal distributions}
\label{sec:appendix_B}
In Section~\ref{sec:emissivities} we wrote the emissivity in the form:

\begin{equation}
    \overline{\varepsilon} = \int \textrm{d}M_{h} \frac{\textrm{d}n(M_{h},z)}{\textrm{d}M_{h}} L (M_{h})
\end{equation}

where we ignore the escape fraction for the moment. As we mentioned in that section, this formula holds if galaxy properties are deterministic functions of halo mass. However, we know that is not the case and for that reason we wrote the general formula for the mean in Eq.~\ref{eq:full_emissivity}. In our case, we assume log-normal distribution for most of the scaling relations (except for the ones relating to the halo abundances) so we can analytically integrate over some of the distributions in Eq.~\ref{eq:full_emissivity}. For simplicity, in this section we reduce the dependencies of luminosity to only $L(M_{h})$ with an appropriate log-normal PDF $p(\log L|M_{h}) = \mathcal{N}(\mu_L,  \sigma_{L})$ without loss of generality. The Eq.~\ref{eq:full_emissivity} becomes:

\begin{align}
\begin{split}
    \overline{\varepsilon} =& \int \textrm{d}M_{h} \frac{\textrm{d}n(M_{h})}{\textrm{d}M_{h}} \int \textrm{d}L \ L \ p(\log_{10} L | \log_{10} M_{h}) \\
    =& \int \textrm{d}M_{h} \frac{\textrm{d}n(M_{h})}{\textrm{d}M_{h}} \int \textrm{d}L \ L \ \frac{1}{\sqrt{2\pi} \log_{e}{10} \ \sigma_{L} L} \exp \left( -\frac{(\log_{10} L - \mu_L)^2}{2 \sigma_{L}^2} \right) 
\end{split}
\end{align}

The integral on the right can be analytically computed:

\begin{equation}
    \overline{\varepsilon} = \int \textrm{d}M_h \frac{\textrm{d}n(M_{h})}{\textrm{d}M_{h}} 10^{\left(\mu_L + \frac{\log_{e}{10} \ \sigma_L^2}{2}\right)}
\label{eq:appendix_integrated}
\end{equation}

Since the halo mass function in general is not an analytic function of halo mass and mean of the luminosity scaling depends on the halo mass, this integral cannot be computed analytically. However we can already gain intuition about the mean looking at the last term of  Eq.~\ref{eq:appendix_integrated}. The factor $10^{\mu_L}$ corresponds to the mean of the $L(M_h)$ scaling relation, i.e. what one would obtain if one did not consider scatter around the mean. The second part, $10^{\frac{\log_{e} 10 \ \sigma_L^2}{2}}$ represents the shift of the mean when integrating over the whole PDF of the distribution. This is proportional to the width of the distribution, indicating that the wider the distribution, the larger the corresponding shift in the mean. This is a general property of asymmetric distributions like the log-normal and has important implications for interpreting means of scaling relations. This is clearly seen in Figures.~\ref{fig:LyC-all},\ref{fig:X-all} and \ref{fig:LW-all} where removing one source of scatter reduces the mean proportionally to the width of the distribution.

If we instead add an additional term that is binomially distributed (e.g. one choice for the escape fraction in Sec.~\ref{sec:escape_fraction}) then the equation becomes:

\begin{align}
\begin{split}
    \overline{\varepsilon} = \int \textrm{d}M_{h} \frac{\textrm{d}n(M_{h})}{\textrm{d}M_{h}} \int \textrm{d}L \ L \ &p(\log_{10} L | \log_{10} M_{h})\times \\& \times \int \textrm{d}f_{\rm esc} f_{\rm esc}{{n}\choose{k}} \mathcal{P}^n(1-\mathcal{P})^k %\\
    %=& \int \textrm{d}M_{h} \frac{\textrm{d}n(M_{h})}{\textrm{d}M_{h}} \int \textrm{d}L \ L \ \frac{1}{\sqrt{2\pi} \ln{10} \ \sigma_{L} L} \exp \left( -\frac{(\log L - \mu_L)^2}{2 \sigma_{L}^2} \right) 
\end{split}
\end{align}
where we have explicitly written out the binomial distribution. In the above, $n=1$, $k=0$ and $\mathcal{P}=\overline{f}_{\rm esc}$ so the distribution trivially becomes:

\begin{align}
\begin{split}
    \overline{\varepsilon} =& \int \textrm{d}M_{h} \frac{\textrm{d}n(M_{h})}{\textrm{d}M_{h}} \int \textrm{d}L \ L \ p(\log_{10} L | \log_{10} M_{h}) \overline{f}_{\rm esc}
\end{split}
\end{align}
Therefore, the mean does not change if scatter is added following a binomial distribution. This is clearly seen in Fig.~\ref{fig:LyC-all}.
\end{document}